\RequirePackage{fix-cm}
\documentclass[smallcondensed]{svjour3}     
\smartqed  

\usepackage{lineno,hyperref}
\modulolinenumbers[5]
\usepackage{graphicx}
\usepackage[dvipsnames]{xcolor}
\usepackage{latexsym}
\usepackage[all]{xy}
\usepackage{mdframed}
\usepackage{multicol} 
\usepackage{framed}
\usepackage{afterpage}
\usepackage{lscape}
\usepackage{subcaption}
\usepackage{setspace}
\usepackage{placeins}
\usepackage{pgfplots}
\usepackage{authblk}
\usepackage{natbib}

\usepackage{lipsum}
\makeatletter
\def\ps@pprintTitle{%
	\let\@oddhead\@empty
	\let\@evenhead\@empty
	\def\@oddfoot{}%
	\let\@evenfoot\@oddfoot}
\makeatother

\newenvironment{ctabbing}%
    {\begin{center}\begin{minipage}{\textwidth}\begin{tabbing}}%
    {\end{tabbing}\end{minipage}\end{center}}

\newcommand{\winners}{\mathcal{W}}
\newcommand{\losers}{\mathcal{L}}
\newcommand{\cand}{\mathcal{C}}
\newcommand{\stand}{\mathcal{S}}
\newcommand{\ballots}{\mathcal{B}}
\newcommand{\election}{\xi}

\newcommand{\elim}{E}
\newcommand{\tally}{t}
\newcommand{\proj}{p}

\newcommand{\true}{\mathit{true}}
\newcommand{\false}{\mathit{false}}
\newcommand{\first}{\mathit{first}}

\newcommand{\ignore}[1]{}
\newcommand{\plusplus}{+\!\!+}

\newcommand{\subfloat}[4][LABEL]{%
	\begin{subfigure}{#3\textwidth}%
	#2%
	\caption{#1}%
	\label{#4}%
	\end{subfigure}%
}

\smartqed
\paperwidth=\dimexpr
1in + \oddsidemargin
+ \textwidth
+ 1in + \oddsidemargin
\relax
\paperheight=\dimexpr
1in + \topmargin
+ \headheight + \headsep
+ \textheight
+ 1in + \topmargin
\relax
\usepackage[pass]{geometry}\relax

\begin{document}
\title{RAIRE: Risk-Limiting Audits for IRV Elections}

\author{Michelle Blom$^1$\thanks{Corresponding author: Michelle Blom, michelle.blom@unimelb.edu.au} \and Peter J. Stuckey$^2$ \and Vanessa J. Teague$^1$}

\institute{$^1$School of Computing and Information Systems \\
The University of Melbourne \\
Parkville, Australia \\
\email{[michelle.blom,vjteague]@unimelb.edu.au}\\
\\
$^2$Faculty of Information Technology \\
Monash University \\
Clayton, Australia \\
\email{peter.stuckey@monash.edu}}

\date{}

\maketitle

\begin{abstract}
Risk-limiting post election audits guarantee a high probability of
correcting incorrect election results, independent of why the result was
incorrect. Ballot-polling audits select cast ballots at random and interpret those ballots
as evidence for and against the reported result, continuing this
process until either they support the reported result, or they fall back to a full manual recount. For elections with digitised scanning and counting of ballots, a ballot-level comparison audit compares randomly selected digital ballots with their paper versions. Discrepancies are referred to as errors, and are used to build evidence against or in support of the reported result. Risk-limiting audits for first-past-the-post elections are well understood, and
used in some US elections.
We define a number of approaches to ballot-polling and ballot-level comparison risk-limiting audits for
Instant Runoff Voting (IRV) elections.
We show that for almost all real elections for which data is available, we can perform a
risk-limiting audit by looking at only a small fraction of the total
ballots (assuming no errors were made in the tallying and distribution of votes).
The techniques presented in this paper represent
 the first practical techniques for conducting risk-limiting audits of IRV elections.
\end{abstract}

\section{Introduction}

Instant Runoff Voting (IRV) is a system of preferential voting in which voters
rank candidates in order of preference. IRV is used for all parliamentary lower
house elections in Australia, parliamentary elections in Fiji and Papua New
Guinea, presidential elections in Ireland and Bosnia/Herzogovinia, and local
elections in numerous locations world-wide, including the UK and United States.
Each voter casts a ballot in which they rank a set of available candidates in
order of preference. In an election with candidates $c_1$, $c_2$, and $c_3$,  a ballot with the ranking [$c_1, c_2, c_3$] expresses a first
preference for candidate $c_1$, a second preference for $c_2$, and a third
for $c_3$. Voters may or may not be required to
number every candidate on the ballot. A ballot may therefore be a partial 
ordering over candidates (e.g., [$c_2, c_1$]).  

The tallying of ballots in an IRV election proceeds by first giving each ballot to its
first ranked candidate. Each candidate has a pile of ballots (called a \textit{tally pile}),
with the number of ballots in their pile called their \textit{tally}. 
 The candidate with the smallest tally (i.e., the candidate with the smallest number of ballots
in their tally pile) is eliminated. The ballots in their tally pile are redistributed to subsequent, less preferred
candidates on the ballot. Elimination proceeds in this fashion, until a single candidate
$w$ remains, who is declared the winner.

The scanning and digitisation of ballots, and the use of automated counting software for computing the outcomes of elections, is becoming more commonplace. 
In light of recent attempts by foreign powers to interfere in electoral processes in the US~\citep{norden2017securing}, there is a growing need for efficient and statistically sound electoral audits. \emph{Risk Limiting Audits} \citep{lindemanStark12} (RLAs) provide strong statistical evidence that the reported outcome of an election is correct, or revert to a manual recount if it is wrong. 
The probability that the audit fails to detect a wrong outcome is bounded by a \textit{risk limit}. An RLA with a risk limit of 1\%, for example, has at most a 1\% chance of failing to detect that a reported election outcome is wrong. In this paper we present several methods for undertaking both ballot-polling and ballot-level comparison RLAs of IRV elections, and compare the auditing effort required by each on a set of real IRV election instances. We show that we can design RLAs for IRV elections that, in general, require only a small fraction of cast ballots to be sampled. 

Risk limiting ballot-polling and ballot-level comparison audits have been developed for first-past-the-post (FPTP) or $k$-winner plurality elections \citep{lindemanEtal12,stark2010super}. In such elections, the $k$ candidates with the most votes are declared winners. A \emph{ballot-polling RLA} of such an election randomly samples the paper ballots cast (or records produced). As each ballot is examined, we update a series of statistics representing hypotheses that any loser actually received more votes than any winner. Once we have seen enough ballots to be confident that all these hypotheses can be rejected, the reported outcome is correct and the audit concludes.  At any point, the audit can fall back to a full manual recount, for example if it is taking too long or has examined a large number of ballots.  The audit is designed so that the probability of concluding with acceptance, when the result is in fact wrong, is at most the risk limit $\alpha$.  \emph{Ballot-level comparison RLAs} are applicable in settings where paper ballots have been scanned and digitised, or a paper-based electronic voting system has been used, producing an index that allows individual electronic ballots to be matched to the paper ballot they represent. Each sampled ballot is compared to its corresponding electronic record. An erroneous ballot is one that does not match its electronic record. These errors are then used to update a similar set of statistics representing hypotheses that the reported election outcomes are actually wrong.     

In this paper we present several methods for undertaking both ballot-polling and ballot-level comparison RLAs of IRV elections. Our ballot-polling RLAs adapt a ballot-polling RLA method (BRAVO), designed for first-past-the-post (FPTP) or $k$-winner plurality elections \citep{lindemanEtal12}, to IRV. Our ballot-level comparison RLAs adapt a comparison-based RLA method (MACRO), similarly designed for FPTP elections \citep{stark2010super}, to IRV.

Of the set of approaches we present for undertaking ballot-polling and ballot-level comparison RLAs of IRV elections, 
the most straightforward views, and audits, each round of an IRV election as a multiple-winner plurality election. This approach has previously been considered (but not evaluated) by \cite{sarwate2013risk}.
 A more efficient method, requiring fewer ballot samples, seeks to prove that the reported winner could not have been eliminated before any other candidate. The former approach is designed to confirm the correctness of each elimination in the IRV counting process, while the latter aims to to confirm \textit{only} that the reported winner of the election is correct. 

The final method we consider uses a custom algorithm, RAIRE, to generate a collection of \textit{assertions} to audit for a given IRV election. These assertions correspond to properties that, if confirmed with a given degree of statistical confidence, confirm that the reported winner is the correct one to the given degree of statistical confidence.  An example of such an assertion is that ``candidate $c$ has a higher tally than candidate $c'$ in the context where candidates $\cand$ have been eliminated''. We show that we can apply BRAVO and MACRO to audit each of these assertions. Our RAIRE algorithm is designed to find the set of such assertions requiring the least estimated auditing effort to prove. 

We compare each of the presented approaches for IRV risk limiting audits  in terms of the number of ballot polls required to audit a number of real election instances with varying risk limits. We experimentally consider audits where all ballots are indeed recorded correctly, and elections where discrepancies exist between paper ballots and their electronic records. Our evaluation compares the relative efficiency of ballot-polling and comparison-based RLAs, in terms of the level of auditing effort required. In the context of FPTP elections, comparison RLAs require fewer ballot samples, in general, to confirm the correctness of an election result.  This is because they can assess, and make use of, the differences between reported and actual individual ballots, a significant source of extra information.

This paper is structured as follows. Related work is described in Section \ref{sec:RelatedWork}. Required background and definitions are presented in Section \ref{sec:Prelim}. Section \ref{sec:Bravo} describes the ballot-polling~\citep{lindemanEtal12} and ballot-level comparison-based RLAs~\cite{stark2010super} upon which our IRV audits are based. We present our IRV ballot-polling and ballot-level comparison RLAs in Sections \ref{sec:OurAudits} and \ref{sec:GeneralRLABPCP}, and evaluate their efficiency in Section \ref{sec:CompResults}.

\section{Related Work}\label{sec:RelatedWork}

Post-election audits are a key measure for increasing both security in our electoral systems, and public confidence in the integrity of our elections \citep{norden2017securing}.
Risk-limiting audits of reported election results against paper ballots (or records) represent the current best-practice for verifying the integrity of an election \citep{rivest2017election}, and a central component of conducting evidence-based elections \citep{lindeman2017evidence}.   

There is a growing literature on the use of risk-limiting audits (RLAs) for auditing the outcome of varying types of election (see e.g. \cite{stark09b,hallEtal09,checkoway2010single,SOBA11,lindemanEtal12,stark2012gentle,sarwate2013risk}). 
RLAs have been applied to a number of plurality (first-past-the-post) elections, including four 2008 elections in California \citep{hallEtal09} and elections in over 50 Colorado counties in 2017 \citep{lindeman2018next}. \cite{stark2014verifiable} present RLAs for D'Hondt (and similar) elections, applicable in a number of European countries such as Norway, Germany, Luxembourg, and Denmark.  General auditing procedures designed to enhance electoral integrity have been outlined by \cite{antonyan2009state}. The BRAVO ballot-polling RLA \citep{lindemanEtal12}, designed for FPTP elections, forms the basis of our IRV ballot-polling audits. The ballot-level comparison RLA presented by \cite{stark2010super} forms the basis of our IRV comparison RLAs. 

A straightforward RLA of an IRV election can be conducted by treating each IRV round as a separate FPTP election. This idea was described by \cite{sarwate2013risk} although not explored in algorithmic detail.  \cite{sarwate2013risk} consider two additional approaches for designing a comparison audit of an IRV election. The first determines whether replacing an erroneous ballot with its correct representation changes the margin of victory of the election.  The second samples $K$ ballots and checks whether the number of erroneous ballots in the sample exceeds a threshold whose value is based on the election's margin of victory.
We demonstrate, however, that we can more efficiently audit an IRV election outcome by simply verifying that the reported winner was not defeated by any other candidate.

An alternative to RLAs are Bayesian audits, proposed by \cite{rivest2012bayesian}. Each ballot is viewed as one of a set of possible ballot types (i.e., the set of all possible rankings over candidates).  In an IRV election that requires voters to rank \textit{all} candidates on a ballot, this set contains $n!$ rankings in a contest with $n$ candidates. In an election with $m$ voters, an election profile is a sequence of $m$ ballot types. 
A Bayesian audit starts with a prior distribution on the set of possible election profiles. This prior could indicate that each profile is as likely to be the true election profile as all others, or that some profiles are more likely than others. As ballots are sampled in a Bayesian audit, they are used to form a posterior probability distribution over the set of possible profiles. Monte Carlo simulation is applied to compute election outcomes across a large number of simulated profiles, drawn from this distribution. The number of simulated outcomes in which the reported winner did not win defines an upset probability. The auditor may terminate the audit when this upset probability falls below a defined threshold. Bayesian audits can, in principle, be applied to any voting system.    
    
The margin of victory (MOV) of the election provides an indication of how many ballots will need to be sampled during a RLA. Elections with a smaller MOV are likely to require more ballots to be sampled in an audit. Automatic methods for computing electoral margins for IRV elections have been presented by \cite{Magrino:irv}, \cite{blom16}, and \cite{beckert2016automatic}. We use prior work for computing the MOV of an IRV election \citep{blom16} as the basis for the RAIRE algorithm. This algorithm is used to build a minimal set of assertions to audit for a given IRV election that, if confirmed, confirm the reported outcome of the election. The set of assertions is minimal in the sense that there exists no other set of assertions that (i) requires fewer estimated ballot polls to audit and (ii) confirms the reported outcome.

\section{Preliminaries}\label{sec:Prelim}

\subsection{First-past-the-post (FPTP)}
 In a single-winner FPTP election, a voter marks a single candidate on their ballot when casting their vote. The candidate who receives the most votes is declared the winner. The BRAVO ballot-polling RLAs \citep{lindemanEtal12}, and MACRO ballot-level comparison RLAs \citep{stark2010super}, are designed for $k$-winner FPTP contests. A voter marks up to $k$  candidates on their ballot, and the $k$ candidates with the highest number of votes are declared winners.  

\subsection{Instant Runoff Voting (IRV)} 

IRV is a form of preferential voting in which voters express a preference ordering over a set of candidates on their ballot. 
The tallying of ballots proceeds by a series of rounds (see
Figure \ref{alg:IRV}) in
which the candidate with the smallest tally (number of ballots in their tally pile) is eliminated,  with the last remaining candidate declared the winner.
All ballots in an eliminated candidate's  tally pile are distributed to the next 
most-preferred eligible candidate in their ranking. A candidate is eligible if they have not yet been eliminated. 

\begin{figure}[t]
\centering
\begin{ctabbing}
xx \= xx \= \kill
Initially, all candidates remain standing (are not eliminated)\\
\textbf{While} there is \textit{more than one} candidate standing \\
\> \textbf{For} every candidate $c$ standing\\
\> \> Tally (count) the ballots in which $c$ is the highest-ranked \\
\>\> candidate of those standing\\
 \> Eliminate the candidate with the smallest tally\\
The winner is the one candidate not eliminated
\end{ctabbing}
\caption{IRV (\textit{a.k.a.} Alternate Vote or Ranked Choice Voting) counting procedure.}
\label{alg:IRV}
\end{figure}

\begin{definition}[IRV election] An IRV election is defined as a tuple $\election = (\cand, \ballots)$
where $\cand$ is the set of available candidates, and $\ballots$ a multiset\footnote{A multiset
allows for the inclusion of duplicate items.} of ballots. Each ballot $b \in
\ballots$ is a sequence of candidates in $\cand$, with no duplicates, listed in
order of preference (most preferred to least preferred).   
\end{definition}

We refer to sequences of candidates $\pi$ in list notation (e.g., $\pi =
[c_1,c_2,c_3]$), and use such sequences to represent both ballots and
the order in which candidates are eliminated. 
We  use the notation $\first(\pi) = \pi(1)$ to denote the first candidate
in a sequence $\pi$.
In each round of vote counting, there are a current set of eliminated candidates $\elim$ and a current set of candidates still standing $\stand = \cand \setminus \elim$. The winner $c_w$  is the last standing candidate.

Each candidate $c \in \cand$ has a \textit{tally pile} of ballots.
Ballots are added to this pile upon the elimination of a candidate $c' \in \cand \setminus c$, and are redistributed upon the elimination of $c$. A candidate's \textit{tally} is equal to the number of ballots in their tally pile. 
We use the concept of \textit{projection} to formally define a candidate's tally at any stage in the IRV counting process.  

\begin{definition}{\textbf{Projection} $\mathbf{\proj_\mathcal{S}(\pi)}$}  We define
  the projection of a sequence $\pi$ onto a set $\mathcal{S}$ as the largest
subsequence of $\pi$ that contains only elements of $\mathcal{S}$. (The elements
  keep their relative order in $\pi$). For example: 

$\proj_{\{c_2,c_3\}}([c_1,c_2,c_4,c_3]) = [c_2,c_3]$ 
 and $\proj_{\{c_2,c_3,c_4,c_5\}}([c_6,c_4,c_7,c_2,c_1]) = [c_4,c_2].$
\label{def:Projection}
\end{definition}

\begin{definition}{\textbf{Tally} $\mathbf{\tally_\stand(c)}$} Given candidates
$\stand \subseteq \cand$ are still standing in an election $\election = (\cand, \ballots)$, the
tally for a candidate $c \in \cand$, denoted $\tally_\stand(c)$, is defined as
the number of ballots $b \in \ballots$ for which $c$ is the most-preferred
candidate of those remaining. Recall that  $\proj_\stand(b)$ denotes the sequence of candidates mentioned in $b$ that are also in $\stand$. Square brackets have been used to denote a multiset.
\begin{equation}
\tally_\stand(c) = ~|~ [b ~|~ b \in \ballots, c = \textit{first}(\proj_\stand(b))] ~|~
\end{equation} 
\label{def:Tally}
\end{definition}   

Several of the auditing methods we propose in this paper make use of a candidate's \textit{primary vote}. 
 
\begin{definition}{\textbf{Primary vote} $\mathbf{f(c)}$}
The \textit{primary vote} of candidate $c \in \cand$, denoted $f(c)$, is the
number of ballots $b \in \ballots$
for which $c$ is ranked highest. \\

\noindent Note that $f(c) = \tally_{\cand}(c)$.

\begin{equation}
f(c) = ~|~ [b ~|~ b \in \ballots, c = \mathit{first}(b) ] ~|~
\label{eqn:pv}
\end{equation}
\end{definition}

\begin{table}[t]
\centering
    \subfloat[]{
      \centering
        \begin{tabular}{cr}
\hline
Ranking & Count \\
\hline
{}[$c_2$, $c_3$] & 4000 \\
{}[$c_1$] &  20000 \\
{}[$c_3$, $c_4$] & 9000 \\
{}[$c_2$, $c_3$, $c_4$] & 6000 \\
{}[$c_4$, $c_1$, $c_2$] &  15000 \\
{}[$c_1$, $c_3$] & 6000 \\
\hline
\end{tabular}}{0.3}{}
    \subfloat[]{
      \centering
        \begin{tabular}{crrr}
\hline
Candidate & Rnd1 & Rnd2 & Rnd3 \\
\hline
{}$c_1$ & 26000  & 26000 & 26000\\
{}$c_2$ &  10000 & 10000 & ---\\
{}$c_3$ & 9000   & --- & ---\\
{}$c_4$ & 15000  & 24000 & 30000 \\
\hline
& $t_{\{c_1,c_2,c_3,c_4\}}$ & $t_{\{c_1,c_2,c_4\}}$ & $t_{\{c_1,c_4\}}$ \\
& & & \\
\end{tabular}}{0.5}{} 
    \caption{An example IRV election, stating (a) the number of ballots cast
with each listed ranking over four candidates, and (b)
 tallies after each round.}
\label{tab:EGIRV}
\end{table}

\begin{example} \label{ex:irv}
Consider the IRV election of Table \ref{tab:EGIRV}. The tallies of 
$c_{1},c_2,c_3,$ and $c_4$, in the $1^{st}$ counting round are 26000, 10000, 9000, and 15000. These tallies represent each candidate's primary vote. Candidate
$c_3$ is eliminated, and 9000 ballots are  distributed to $c_4$, who now has a tally 
of 24000. Candidate $c_2$, on 10000 votes, is eliminated next with 6000 of their ballots
given to $c_4$ (the remainder have no subsequent preferences). Candidates $c_1$ and $c_4$ remain with tallies of 26000 and 30000. Candidate  $c_1$ is eliminated and $c_4$ elected.
\end{example}

The aim of a RLA is to either gain evidence that the reported results are correct (to some risk limit $\alpha$)
or to correct an incorrect result by falling back to a manual recount. We distinguish between reported results 
(tallies and counts based on digital ballot records) and the \emph{actual} results which should have been
calculated, as represented in the paper ballot records. We use a tilde $\sim$ to refer to these actual results. For example, $f(c)$ is the
recorded primary vote for candidate $c$ and $\tilde{f}(c)$ is the actual
primary vote for the candidate. Similarly, we refer to reported ballots with $\ballots$ and actual ballots with $\tilde{\ballots}$.

\section{Risk-limiting audits for FPTP}\label{sec:Bravo}

In this section, we describe the BRAVO and MACRO ballot-polling and ballot-level comparison RLAs for $k$-winner 
FPTP elections. These methods form the building blocks for our IRV RLAs (Sections  \ref{sec:OurAudits} and \ref{sec:GeneralRLABPCP}). In a $k$-winner FPTP election, the $k$ candidates with the greatest tallies are elected. All winners are elected simultaneously and there is no
transfer of ballots.  Given a set of $\cand$ candidates ($|\cand| = n)$ there
will be a set of $\winners$ \emph{winners} ($|\winners| = k$) and $\losers$ \emph{losers} ($|\losers|
= n - k$). Both BRAVO and MACRO are applicable in elections where each ballot may express a vote for one or more candidates. We describe these methods in the context where each ballot expresses a vote for a single candidate, and where ballots are sampled one at a time. This is the setting in which they will be simulated in our evaluation. In practice, ballots are drawn in batches during an audit, with an estimate of required auditing effort used to set the size of the initial sample. It would be straightforward to adapt our IRV audits to operate in this setting.    

Throughout this section we use the notation $t(c)$ ($\tilde{t}(c)$) to denote the reported (actual) number of ballots expressing a vote for candidate $c$.  
 
\subsection{BRAVO ballot-polling RLAs}

Figure \ref{fig:bravo} outlines the steps involved in a BRAVO ballot-polling RLA for a $k$-winner FPTP election \citep{lindemanEtal12} .  A BRAVO audit independently tests $k(n-k)$ null
hypotheses $\{ \tilde{t}(w) \leq \tilde{t}(l) \}$, one for each winner/loser pair. For winner $w$ and loser $l$, the hypothesis $\{ \tilde{t}(w) \leq \tilde{t}(l) \}$ states that $l$ has amassed at least as many votes as $w$ (i.e., that $w$ does not actually beat $l$). 

A statistic for each of these hypotheses, $T_{wl}$, is maintained and updated when a ballot is drawn that expresses a vote for $w$ or $l$. Each $T_{wl}$ is initialised to 1, as shown in Figure \ref{fig:bravo}. When a ballot expressing a vote for winner $w$  is drawn, the $T_{wl}$ statistics for $w$ and each $l \in \losers$ are increased. When a ballot expressing a vote for loser $l$ is drawn, the $T_{wl}$ statistics for $l$ and each $w \in \winners$ are decreased. We reject a given hypothesis once $T_{wl}$ exceeds a given threshold, equal to $1/\alpha$ where $\alpha$ is the risk limit of the audit. We can estimate for each hypothesis the number of
sampled ballots we expect will be required to reject
it, assuming the announced election counts are accurate.

Let $p_c$ be the proportion of recorded ballots $\ballots$ expressing a vote for candidate $c$,
i.e. $p_c = t(c) / |\ballots|$. Of the ballots expressing a vote for either candidate $w$
or $l$, the proportion of these expressing a vote for $w$ is denoted $s_{wl}$.
\begin{equation}
s_{wl} = \frac{p_{w}}{p_w + p_l}
\end{equation}

\cite{lindemanEtal12} define the \emph{Average Sample Number} (ASN) for a BRAVO audit with risk limit $\alpha$ -- 
 the number of ballots we expect we will need to sample to reject 
the null hypothesis $\{ \tilde{t}(w) \leq \tilde{t}(l) \}$ -- as follows.

\begin{equation}
  ASN \simeq \frac{ln(1/\alpha) + 0.5 ln(2 s_{wl})}{(p_w\, ln(2s_{wl}) + p_l\, ln(2 -
    2s_{wl}))}
\label{eqn:ASNBP}
\end{equation}

To confirm the reported outcome of an election, BRAVO must reject $\winners \times \losers$ null hypotheses. The expected number of ballot samples required to do this is equal to the maximum of the ASNs for each hypothesis. 

\begin{figure}[!t]
\centering
\begin{minipage}{\textwidth}
  \begin{tabbing}
    xxx \= xx \= xx \= xx \= xx \= xx \= \kill
    \textsf{bravo}($\tilde{\ballots}$, $\winners$, $\losers$, $\alpha$, $M$) \\
    \> $\triangleright$ Initialise statistic $T_{wl}$ for each winner/loser pair \\[2pt]
    1 \> \textbf{for}($w \in \winners, l \in \losers$) \\
    2 \> \> $T_{wl}$ := $1$ \\
    3 \> \> $s_{wl}$ := $t(w) / (t(w) + t(l))$ \\[5pt]
    4 \> $H$ := $\winners \times \losers$ \\[5pt]
    \> $\triangleright$ Initialise counter $m$ recording number of ballots sampled thus far \\[2pt]
    5 \> $m$ := 0 \\[5pt]
    \> $\triangleright$ We randomly draw a ballot, updating our statistics $T_{wl}$ with each  \\
    \> ballot drawn, until a maximum number of ballots $M$ have been drawn,  \\
    \> or we have rejected all $|H|$ hypotheses \\[2pt]
    6 \> \textbf{while}($m < M \wedge H \neq \emptyset$) \\
    7 \> \> Randomly draw a ballot $b$ from $\tilde{\ballots}$ \\
    8 \> \> $m$ := $m+1$ \\[5pt]
    9 \> \> \textbf{if}($b$ expresses a vote for a winner $w' \in \winners$) \\
    10 \> \> \> \textbf{for}($(w',l) \in H$) \\
    11 \> \> \> \> $T_{w'l}$ := $T_{w'l} \times 2 s_{w'l}$ \\
    12 \> \> \> \> \textbf{if}($T_{w'l} \geq 1/\alpha$) \\
    13 \> \> \> \> \> $\triangleright$ We reject the null hypothesis $\{ \tilde{t}(w') \leq \tilde{t}(l) \}$  \\
    14 \> \> \> \> \> $H$ = $H - \{ (w',l) \}$ \\[5pt]
    15 \> \> \textbf{else if}($b$ expresses a vote for a loser $l' \in \losers$) \\
    16 \> \> \> \textbf{for}($(w,l') \in H$) \\
    17 \> \> \> \> $T_{wl'}$ := $T_{wl'} \times 2(1 - s_{wl'})$ \\[5pt]
    18 \> \textbf{if}($H = \emptyset$) \\
    \> \> $\triangleright$ The reported outcome stands \\
    19 \> \> \textbf{return} $\true$ \\
    20 \> \textbf{else}  \\
    \> \> $\triangleright$ A full recount is required \\
    21 \> \> \textbf{return} $\false$ 
  \end{tabbing}
\end{minipage}
\caption{A BRAVO RLA of a FPTP
  election with actual ballots $\tilde{\ballots}$, declared winners $\winners$ and
  losers $\losers$, risk limit $\alpha$, and limit on ballots
  checked $M$. }
  \label{fig:bravo}
\end{figure}

\begin{example} \label{ex:first}
  Consider the first round of the IRV election of
  Example~\ref{ex:irv}. If we view this first round as a FPTP election with winners $c_1$, $c_2$, and $c_4$, and loser $c_3$, the null hypotheses we need to reject are
  $\tilde{f}(c_1) \leq \tilde{f}(c_3)$,  $\tilde{f}(c_2) \leq \tilde{f}(c_3)$,
  $\tilde{t}(f_4) \leq \tilde{f}(c_3)$. Recall that $f(c)$ and $\tilde{f}(c)$ represent candidate $c$'s reported, and actual, tallies in the first round of an IRV election.  We calculate $p_1 = 26000/60000$,
  $p_2 = 10000/60000$, $p_3 = 9000/60000$, $p_4 = 15000/60000$ and
  $s_{13} = 26000/35000$, $s_{23} = 10000/19000$, and $s_{43} = 15000/24000$.
  The ASN for rejecting each hypothesis using BRAVO, assuming $\alpha
  = 0.05$, is 44.5, 6885, and 246 ballots respectively. We expect we will need to sample 6885 ballots during the audit as a whole.
\end{example}

The underlying statistical principles of the BRAVO RLA are described by \cite{lindemanEtal12}. 

\subsection{MACRO ballot-level comparison RLAs for FPTP}\label{sec:CompRLA}

\cite{stark2010super} presents a method for conducting a ballot-level comparison RLA of a collection of FPTP contests simultaneously. We refer to this audit as MACRO. In this section, we describe MACRO in the context of a single race, where each ballot records a vote for a single candidate. This audit randomly samples paper ballots from the set $\tilde{\ballots}$ and finds the matching electronic records  in $\ballots$. For each such ballot, we compare the paper $\tilde{b}$ and recorded $b$ representations. We assess the difference in these representations in terms of the extent to which any error has overstated a pairwise margin between a winner and loser. The procedure followed by MACRO is shown in Figure \ref{fig:macro}. 

For each sampled ballot, we compute its maximum across-contest relative overstatement (MACRO). In the single-contest setting, this statistic could be denoted the maximum relative overstatement.  In an election with winners $\winners$, and losers $\losers$, the MACRO for a ballot $b$ is defined by \cite{stark2009auditing} as:
\begin{equation}
e_b = \max_{w \in \winners, l \in \losers} (v_{bw} - a_{bw} - v_{bl} + a_{bl}) / V_{wl}
\label{eqn:ep}
\end{equation}  
where: $v_{bc} \in \{0,1\}$ is 1 if $b$ is a recorded vote for candidate $c$, and 0 otherwise; $a_{bc} \in \{0,1\}$ is 1 if $\tilde{b}$ is an actual vote for candidate $c$; and $V_{ij}$ the pairwise margin (difference in recorded tallies) between candidates $i$ and $j$. For a given recorded ballot $b$, Equation \ref{eqn:ep} computes the maximum extent to which any discrepancy between $b$ and $\tilde{b}$ has overestimated a pairwise margin between a winner $w$ and loser $l$ pair. For example, consider the situation in which a paper ballot $\tilde{b}$ expresses a vote for loser $l$ ($a_{bl} = 1$). Its electronic record expresses a vote for winner $w$ ($v_{bw} = 1$, $v_{bl} = 0$ and $a_{bw} = 0$). This discrepancy has overestimated the pairwise margin between $w$ and $l$ by 2 votes.  

Recall that a BRAVO audit maintains a statistic $T_{wl}$ for each winner/loser pair. These statistics are updated as ballots are sampled, and are used as a stopping criterion for the audit. In a MACRO audit, a different statistic is maintained -- a running Kaplan-Markov MACRO P-value ($P_{KM}$). As each ballot $b$ is sampled, we multiply $P_{KM}$ as shown below \citep{stark2010super}:
\begin{equation}
P_{KM} = P_{KM} \times \frac{1 - 1/U}{1 - \frac{e_b}{2\gamma/V_{min}}}
\end{equation}
where: $V_{min}$ is the smallest recorded margin between a winning and losing candidate (the difference in their tallies);  $\gamma \geq 1$ an ``error inflation factor'' (an operational parameter for the audit, described below); $U = (2\gamma|\ballots|)/V_{min}$; and $e_b$ is defined as per Equation~\ref{eqn:ep}. We continue to sample ballots until either a maximum number $M$ of ballots have been checked (indicating that a full recount is required), or $P_{KM}$ falls below the risk limit $\alpha$ of the audit.

\begin{figure}[!t]
\centering
\begin{minipage}{\textwidth}
  \begin{tabbing}
    xxx \= xx \= xx \= xx \= xx \= xx \= \kill
    \textsf{macro}($\ballots$, $\tilde{\ballots}$, $\winners$, $\losers$, $M$, $\alpha$, $\gamma$) \\
    \> $\triangleright$ Compute the smallest pairwise margin between a winner and loser \\
	1 \> $V_{min}$ := $\min_{w \in \winners, l \in \losers} t(w) - t(l)$ \\[2pt]
    2 \> $\mu$ := $\frac{V_{min}}{| \ballots |}$ \\[2pt]
    3 \> $U$ := $2\gamma/\mu$ \\[5pt]
    \> $\triangleright$ Initialise running Kaplan-Markov MACRO P-value\\[2pt]
	4 \> $P_{KM}$ := $1$ \\[5pt]
    \> $\triangleright$ Initialise counter $m$ recording number of ballots sampled thus far \\[2pt]
	5 \> $m$ := $1$ \\
    \> \textbf{repeat} \\
    6 \> \> Randomly draw a ballot $\tilde{b} \in \tilde{\ballots}$ and matching reported ballot $b \in \ballots$ \\[5pt]
	7 \> \> For each candidate $c$, define $v_{bc} = 1$ if $b$ expresses a vote for \\
    \> \> $\quad$ candidate $c$ and $a_{bc} = 1$ if $\tilde{b}$ expresses a vote for $c$, \\
    \> \> $\quad$ and $v_{bc} = a_{bc} = 0$ otherwise\\[5pt]
    \> \> $\triangleright$ Compute the maximum relative overstatement caused by any \\
    \> \> discrepancy between $b$ and $\tilde{b}$ \\[2pt]
	8 \> \> $e_b$ := $\max_{w \in \winners, l \in \losers} \frac{v_{b w} - a_{b w} - v_{b l} + a_{b l}}{t(w) - t(l)}$ \\[5pt]
	9 \>\> $P_{KM} = P_{KM} \times \frac{1 - 1/U}{1 - \frac{e_b}{2\gamma/V_{min}}}$ \\[5pt]
	10 \>\> \textbf{if}($P_{KM} \leq \alpha$) \\
    \> \>\> $\triangleright$ The reported outcome stands \\
	11 \>\>\> \textbf{return} $\true$ \\[5pt]
	12 \>\> $m$ := $m + 1$ \\
    13 \> \textbf{until}($m \geq M$) \\[5pt]
	\> $\triangleright$ A full recount is required \\
    14 \> \textbf{return} $\false$ 
  \end{tabbing}
\end{minipage}
\caption{MACRO ballot-level comparison RLA of a FPTP election with actual ballots $\tilde{\ballots}$, reported ballots $\ballots$, declared winners $\winners$ and losers $\losers$, inflation factor $\gamma$, risk limit $\alpha$ and sampling limit $M$. }
  \label{fig:macro}
\end{figure}

When applied in practice, two parameters ($\gamma \geq 1$ and $\lambda < 1$) influence the size of the initial sample of ballots drawn in a MACRO audit, the expected additional auditing effort required when certain types of discrepancies arise during the audit (i.e., 2-vote overstatements), and present additional stopping conditions. The ``error tolerance'' $\lambda$ is relevant only when a batch of ballots is drawn as the initial sample, as opposed to a single ballot, and consequently we set $\lambda = 0$ when applying MACRO in our IRV audits. We experiment with varying values for $\gamma$ when simulating our IRV audits. 

Given an overall risk limit $\alpha$, we can estimate the number of ballots that must be sampled by a MACRO audit under the assumption that no errors are present in the electronic ballot records. We reuse the terminology of ballot polling audits, and call this number of ballots the \emph{Average Sample Number} (ASN) for the audit. Given an election with reported ballots $\ballots$, the ASN for a MACRO audit, with risk limit $\alpha$, is defined by \cite{stark2010super} as:
\begin{equation}
  ASN \simeq -ln(\alpha) \, U
\label{eqn:ASNCP}
\end{equation}
where $U$ is defined as above. Equation \ref{eqn:ASNCP} is derived following Steps 4 and 6 on page 8 of \cite{stark2010super} with $\lambda = 0$.
\begin{example}

Consider the first round of the IRV election of Example~\ref{ex:irv}, viewed as a FPTP election with winners $c_1$, $c_2$, and $c_4$, and loser $c_3$. This election can be audited by a single application of MACRO (Figure \ref{fig:macro}). The tallies for each candidate are shown in Table \ref{tab:EGIRV}, column two. The margins between each winner-loser pair in this first round election are $V_{c_1,c_3} = 17000$, $V_{c_2,c_3} = 1000$, and $V_{c_4,c_3} = 6000$. The smallest winner-loser margin $V_{min}$ is 1000. Using the formula in Equation \ref{eqn:ASNCP}, with $\alpha = 0.05$ and $\gamma = 1.1$, the expected number of ballot checks required by MACRO is 395.4, with $U = (2\gamma|\ballots|)/V_{min} = 132$.

When auditing this first-round election, the algorithm of Figure \ref{fig:macro} randomly draws a paper ballot $\tilde{b} \in \tilde{\ballots}$ and compares it to its electronic record $b \in \ballots$. If $\tilde{b}$ and $b$ match, the computed error $e_b$ is equal to 0. Consider the situation in which a paper ballot with ranking $\tilde{b} =$ [$c_3$, $c_4$] has been recorded as a $b =$ [$c_2$, $c_3$, $c_4$] ballot, with the election profile listed in Example~\ref{ex:irv} representing reported counts. To determine the impact of this erroneous recorded ballot, we look at each winner $w$ and loser $l$ pair ($c_1$,$c_3$), ($c_2$, $c_3$), and ($c_4$, $c_3$). For each winner-loser pair, we compute, and take the maximum of, the expression:
\begin{equation}
\frac{v_{bw} - a_{bw} - v_{bl} + a_{bl}}{f(w) - f(l)}
\end{equation}
For ($c_1$, $c_3$) and ($c_4$, $c_3$) the numerator in the above expression is equal to 1 -- the error in the reported ballot overestimated the margin between these winners and the loser by 1 vote. For pair ($c_2$, $c_3$), the numerator in the above expression is equal to 2 -- the error in the reported ballot overestimated the margin between $c_2$ and $c_3$ by 2 votes. For this ballot, $e_b = 2\times10^{-3}$. 
\label{ex:CRLAfirst}
\end{example}

The underlying statistical principles of the MACRO audit are described by \cite{stark2010super}.

\section{Risk-Limiting Audits for IRV Elections}\label{sec:OurAudits}

In this section we present four different approaches for conducting a ballot-polling or ballot-level comparison RLA for an IRV election. The first method audits the entire elimination order, ensuring that every step in the IRV election was correct (with some confidence). The second method simplifies the auditing task in settings where we can eliminate multiple candidates in a single round. The third method seeks to examine only whether the eventual winner was the correct one. The fourth approach is a general algorithm, RAIRE, for finding efficient ballot-polling and ballot-level comparison RLAs for IRV elections.

Each of these involves auditing simultaneously a collection $\mathcal{A}$ of different `assertions' or hypotheses, whose conjunction is what we actually want to check.  In the first case, we are interested in checking the complete elimination order; in later methods we are interested in a collection of assertions which, taken together, imply that the announced winner truly won. Each of these assertions represents a hypothesis that can be checked with a single application of a BRAVO or MACRO audit -- for example, that candidate $c$ has a higher tally than $c'$ when candidates $\stand$ are still standing, and $\cand \setminus \stand$ have been eliminated.  Each of these individual audits is conducted to the same risk limit $\alpha$.  If at any point, any of the audits fail to reach a positive conclusion, we manually recount the whole election.  It is easy to see that this process constitutes a valid risk-limiting audit to risk limit $\alpha$ of the election result, assuming that our collection of chosen assertions does indeed imply that that candidate won.  Suppose that the announced election outcome is actually wrong.  Then at least one assertion in $\mathcal{A}$ must be false.  The individual audit of that assertion will therefore go to a full manual recount with a probability of at least $1-\alpha$, at which point we hand count the whole election.

In the following sections we outline how each of the four approaches for auditing an IRV election, described above, can be conducted as ballot-polling or ballot-level comparison RLAs. 

\begin{figure}[!thp]
\centering
\begin{minipage}{\textwidth}
  \begin{tabbing}
    xxx \= xx \= xx \= xx \= xx \= xx \= \kill
    \textsf{irv-bravo}($\tilde{\ballots}$, $\pi$, $\alpha$, $M$) \\
    1 \> $H$ := $\emptyset$ \\[5pt]
    \> $\triangleright$ Build a set of hypotheses to check by interpreting each round \\
    \> as a $k$-winner FPTP election with a single loser $l$ (the eliminated \\
    \> candidate) and a set of winners $C_l \setminus \{l\}$ \\[5pt]
    2 \> \textbf{for}($i \in 1 .. |\pi|-1$) \\
    \> \> $\triangleright$ $\pi(i)$ denotes the candidate in position $i$ of elimination order $\pi$ \\
    3 \> \> $l$ := $\pi(i)$ \\[5pt]
    \>\> $\triangleright$ $C_l$ denotes the set of candidates still standing in round $i$\\ 
    4 \> \> $C_l$ := $\{ \pi(i), \pi(i+1), \ldots, \pi(|\pi|)\}$ \\[5pt]
    \> \> $\triangleright$ Candidates in positions $i+1$ to $|\pi|$ are  viewed as the winners of a \\
    \> \> FPTP election with one loser, candidate $l$. For each winner $w$/loser $l$ \\
    \> \> pair we want to confirm the hypothesis that $w$ does in fact beat $l$ in \\
    \> \> this election. We do so by rejecting or disproving the null hypothesis \\
    \> \> that $\tally_{C_l}(w) \leq \tally_{C_l}(l)$ \\[5pt] 
    5 \> \> \textbf{for}($j \in i+1 .. |\pi|$) \\
    6 \> \> \> $w$ := $\pi(j)$ \\
    7 \> \> \> $T_{wl}$ := $1$ \\
    8 \> \> \> $s_{wl}$ := $\tally_{C_l}(w) / (\tally_{C_l}(w) + \tally_{C_l}(l))$ \\
    9 \> \> \> $H$ := $H \cup \{(w,l)\}$ \\
    \> $\triangleright$ Initialise counter $m$ recording number of ballots sampled thus far \\[2pt]
    10 \> $m$ := 0 \\[5pt]
    \> $\triangleright$ We randomly draw a ballot, updating our statistics $T_{wl}$ with each  \\
    \> ballot drawn, until a maximum number of ballots $M$ have been drawn,  \\
    \> or we have rejected all $|H|$ hypotheses \\[2pt]
    11 \> \textbf{while}($m < M \wedge H \neq \emptyset$) \\
    12 \> \> randomly draw ballot $b$ from $\tilde{\ballots}$  \\
    13 \> \> $m$ := $m+1$ \\
    14 \> \> \textbf{for}($(w,l) \in H$) \\
    \>\>\>   $\triangleright$ if $b$ expresses a vote for $w$, we increase $T_{wl}$ \\[2pt]
    15 \> \> \> \textbf{if}($w = \first(\proj_{C_l}(b))$) \\
    16 \> \> \> \> $T_{wl}$ := $T_{wl} \times 2 s_{wl}$ \\
    17 \> \> \> \> \textbf{if}($T_{wl} \geq 1/\alpha$) \\
    \> \> \> \> \> $\triangleright$ We reject the null hypothesis that $\tally_{C_l}(w) \leq \tally_{C_l}(l)$ \\[2pt]
    18 \> \> \> \> \> $H$ = $H - \{ (w,l) \}$ \\[5pt]
    \>\>\> $\triangleright$ if $b$ expresses a vote for $l$, we decrease $T_{wl}$ \\[2pt]
    19 \> \> \> \textbf{else if}($l = \first(\proj_{C_l}(b))$) \\
    20 \> \> \> \> $T_{wl}$ := $T_{wl} \times 2(1 - s_{wl})$ \\
    21 \> \textbf{if}($H = \emptyset$) \\
    \> \> $\triangleright$ The reported outcome stands \\
    22 \> \> \textbf{return} $\true$ \\
    23 \> \textbf{else}  \\
    \> \> $\triangleright$ A full recount is required \\
    24 \> \> \textbf{return} $\false$ 
  \end{tabbing}
\end{minipage}
\caption{A risk-limiting ballot-polling RLA of an
  IRV election with actual ballots $\tilde{\ballots}$, order of elimination $\pi$,
risk limit $\alpha$ and limit on ballots checked $M$. }
  \label{fig:irvbravo}
\end{figure}

\subsection{Auditing a particular elimination order}  \label{sec:particularElim}
 
The simplest approach to applying risk limiting auditing to
IRV is to consider the IRV election as a number of simultaneous FPTP
elections, one for each IRV round.
This was suggested by \cite{sarwate2013risk}, but not explored algorithmically.
Note that this may perform much more auditing than required, since it
verifies more than whether the eventual winner was correct,
but that every step in the IRV election was correct (with some confidence).

Given an election $\ballots$ of $n$ candidates $\cand$, we define the computed elimination order as $\pi$ $=$ $[c_1, c_2, \ldots, c_{n-1}, c_n]$ where $c_1$ is
the first eliminated candidate, $c_2$ the second, etc, and $c_n$ the
eventual winner. 

We treat each IRV round as a FPTP election. In round $i$, we have a set of winning candidates ($C_w$, the candidates that are still standing after round $i$) and a single losing candidate ($l = c_i$, the candidate eliminated in round $i$). More formally, the set of candidates in the round $i$ FPTP election is $C_{l} = \{ c_{j} ~|~ i \leq j \leq n\}$. Each candidate $c \in C_l$ has a recorded tally of $t_{C_{l}}(c)$. The loser of this election is $l = c_i$ and the set of $n - i$ winners denoted by $C_l \setminus \{l\}$.   

\subsubsection{Auditing the elimination order by ballot-polling} 
We can audit all these FPTP elections simultaneously by considering
all the null hypotheses that would violate the announced result.
These are $\{ \tilde{\tally}_{C_l}(c) \leq \tilde{\tally}_{C_l}(c_l) ~|~ 1 \leq i \leq n-1, l = c_i, c \in C_i
\setminus \{ l \} \}$ -- i.e., that the eliminated candidate in each round $i$ does not have fewer votes than each of the other candidates that are still standing in that round. We represent these hypotheses by a pair $(w,l)$ of
winner $w = c$, and loser $l = c_i$.
The statistic maintained for this test is $T_{wl}$.
Each loser loses in only one round so there is no ambiguity.

The overall audit proceeds as shown in Figure~\ref{fig:irvbravo}.  The set of hypotheses
$H$ are  pairs $(w,l)$ of winner $w$ and loser $l$,
and are interpreted as a
hypothesis for the FPTP election corresponding to the round where $l$ was
eliminated.  The calculation of the expected ratio of votes
$s_{wl}$ must be made using the tallies from this round, and we must consider every ballot to 
see how it is interesting for that particular hypothesis. As each ballot $b$ is drawn, we
interpret it as a vote for candidate $c$ if $c = first(p_{C_l}(b))$. In this case, 
candidate $c$ is the  first ranked candidate of those still standing in this round.
  A ballot that is exhausted
after $k$ rounds, for example, will not play a role when determining the statistics for later
round hypotheses.

\begin{example} Consider the IRV election of Example~\ref{ex:irv}.
  The null hypotheses we need to reject are
    $\tilde{f}(c_1) \leq \tilde{f}(c_3)$,  $\tilde{f}(c_2) \leq \tilde{f}(c_3)$,
  and $\tilde{f}(c_4) \leq \tilde{f}(c_3)$
  from the first round election,
  $\tilde{\tally}_{\{c_1,c_2,c_4\}}(c_1) \leq
  \tilde{\tally}_{\{c_1,c_2,c_4\}}(c_2)$ and
  $\tilde{\tally}_{\{c_1,c_2,c_4\}}(c_3) \leq
  \tilde{\tally}_{\{c_1,c_2,c_4\}}(c_2)$
  from the second and
  $\tilde{\tally}_{\{c_1,c_4\}}(c_4) \leq
  \tilde{\tally}_{\{c_1,c_4\}}(c_1)$
  from the final round. 
  Assuming $\alpha = 0.05$ the ASNs for the first round are the same as calculated in
  Example~\ref{ex:first}.
  The ASNs required to disprove the two stated null hypotheses for the second round election are 51.8 and 64.0. The ASN required to disprove the final round null hypothesis is 1186. The ASN of the overall audit is the maximum of the ASNs required to disprove all null hypotheses, across each round. For this election, this ASN is 6885.
\end{example}

The weakness of this naive approach is that inconsequential earlier
elimination rounds can be difficult to audit even if they are irrelevant to
the winner.
\begin{example}\label{ex:bad}
 Consider an election with candidates $c_1, c_2, c_3, c_4, c_5$
and ballots  $[c_1] :10000$, $[c_2]: 6000$, $[c_3,c_2]:
3000$, $[c_3,c_1]: 2000$, $[c_4]: 500$,
$[c_5]: 499$. The elimination order is $[c_5,c_4,c_3,c_2,c_1]$.
Given $\alpha = 0.05$, rejecting the null hypothesis that $c_5$ beat $c_4$ in the first round
gives an ASN of \mbox{13,165,239} indicating a full recount is required.
But it is irrelevant to the election result.
\end{example}

\subsubsection{Auditing the elimination order by a comparison audit}\label{sec:particularElimCP}

Each of these FPTP elections can also be audited via a single application of MACRO (Figure \ref{fig:macro}) with $\winners = C_l \setminus \{l\}$, $\losers = \{l\}$, and appropriate instantiations of the risk limit $\alpha$ and inflator $\gamma$ parameters. As in the ballot-polling context, we can audit each of these FPTP elections simultaneously. In contrast to the ballot-polling audit, we need only perform a single comparison RLA (using MACRO) for each IRV round. Our ballot-polling audit applies BRAVO to each of a series of hypotheses in each round (one for each winner-loser pair). 

\begin{example} Consider again the IRV election shown in Example~\ref{ex:irv}. To audit the entire elimination order with a comparison audit, we treat each IRV round as a FPTP election and run a MACRO audit. Assuming $\alpha = 0.05$, and $\gamma = 1.1$, the expected number of ballot checks required by MACRO is the same as that calculated in Example~\ref{ex:CRLAfirst}.
For the remaining two IRV rounds, the ASNs required by MACRO are 28.2 and 98.9. The ASN of the overall audit is the maximum of the ASNs required by MACRO in each round. For this election, this ASN is 395.4. In this case, auditing the entire elimination order by a comparison audit is likely to be more efficient than a corresponding ballot-polling audit.
\end{example}

\subsection{Simultaneous elimination}\label{sec:Batch}

It is common in IRV elections to eliminate multiple candidates in a single
round if it can be shown that the order of elimination cannot affect later
rounds.  Given an elimination order $\pi$ we can simultaneously eliminate
candidates $E = \{ \pi(i) .. \pi(i+k)\}$ if the sum of tallies of these candidates is
less than the tally of the next candidate to be eliminated, $\pi(i+k+1)$. Let $C = \{ \pi(i),
\pi(i+1), \ldots \pi(k), \pi(k+1), \ldots \pi(n)\}$ be the set of candidates
standing after the first $i-1$ have been eliminated.
We can simultaneously eliminate $E$ if:
\begin{equation}
  \tally_C(c) > \sum_{c' \in E} \tally_C(c') 
  \quad \forall c \in C \setminus E
\end{equation}
This is because no matter which order the candidates in $E$ are eliminated
no candidate could ever garner a tally greater than one of the candidates in
 $C \setminus E$. Hence they will all be eliminated in any case.
As the remainder of the election only depends on the set of
eliminated candidates and not their order, the simultaneous elimination can
have no effect on later rounds of the election.

We can take advantage of  simultaneous elimination when auditing by considering
all the simultaneously eliminated candidates $E$ as a single loser $l$. Like the audit of a particular elimination sequence, we are proving a stronger result than necessary, {\it i.e.} that a particular sequence of (possibly multiple) eliminations is valid. Utilising simultaneous elimination, however, often results in a much lower ASN, though not necessarily: sometimes
the combined total of first preferences in $E$ is very close to the next
tally, so a lot of auditing is required.  It may be better to audit each
elimination individually in this case.
It is possible to compute the ASN for each approach and choose the method that requires the least auditing, assuming the outcome is correct.

\subsubsection{Simultaneous elimination by ballot-polling}

We want to reject hypotheses $\tilde{\tally}_{C}(w) \leq \tilde{\tally}_{C}(l)$
for each $w \in C \setminus E$.  The statistic $T_{wl}$ in this case is increased when we draw a
ballot where $w$ is the highest-ranked of remaining candidates $C$, and decreased
when we draw a ballot where $c' \in E$ is the highest-ranked of remaining candidates
$C$.

The elimination of all these null hypotheses is sufficient to prove that the multiple elimination is correct.  This can then be combined with the audit of the rest of the elimination sequence, as described in Section~\ref{sec:particularElim}, to test whether the  election's announced winner is correct. 

\begin{example}
  Consider the election in Example~\ref{ex:bad}.
  We can simultaneously eliminate the candidates $E = \{ c_5, c_4\}$
  since the sum of their tallies $499 + 500 < 5000$ which is the lowest
  tally of the other candidates ($c_1$, $c_2$, and $c_3$).  If we do this the difficult first round
  elimination auditing disappears.  This shows the benefit of simultaneous
  elimination.  The ASNs required for the joint elimination of $E$ are
  17.0, 36.2 and 49.1 ballots, as opposed to requiring a full recount. 

  After this simultaneous elimination, the tallies for the three candidate election
  $\{c_1,c_2,c_3\}$ are $c_1 : 10000$, $c_2 : 6000$ and $c_3: 5000$.  
  The ASNs to reject the hypotheses $\tilde{\tally}_{\cand}(c_1) \leq 
  \tilde{\tally}_{\cand}(c_3)$ and $\tilde{\tally}_{\cand}(c_2) \leq 
  \tilde{\tally}_{\cand}(c_3)$ are 77.6 and 1402 respectively.
  
  We could also simultaneously eliminate  $\{c_5,c_4,c_3\}$
  since the sum of their tallies $499 + 500 + 5000 < 6000$ which is the
  lowest tally of the other candidate (that of $c_2$). But this will lead to a very difficult
  hypothesis to reject, $\tilde{\tally}_{\cand}(c_2) \leq 
  \tilde{\tally}_{\cand}(\{c_5,c_4,c_3\})$ since the tallies are almost identical!
  The ASN is 158,156,493!
  This illustrates that simultaneous elimination may not always be beneficial.
\end{example}

\subsubsection{Simultaneous elimination by a ballot-level comparison RLA}
 As in the ballot-polling context, we treat any simultaneously eliminated candidates $E$ as a single loser $l$, eliminated in a single round $i$. We treat each round as a FPTP election, audited via a single application of MACRO.

\begin{example}
Consider again the election in Example~\ref{ex:bad}, in the setting where we simultaneously eliminate candidates $E = \{ c_5, c_4\}$ in the first round. When viewed as a single losing candidate $l$, the winner-loser pairwise margins in this first round FPTP election are $V_{c_1,l} = 9001$, $V_{c_2,l} = 5001$, and $V_{c_3,l} = 4001$. Assuming $\alpha = 0.05$ and $\gamma = 1.1$, the expected number of ballot checks required by MACRO to audit this first round FPTP election is 36.2. In the second round FPTP election, candidates $c_1$, $c_2$ and $c_3$ remain with winners $\{c_1, c_2\}$ and loser $c_3$. The winner-loser pairwise margins in this election are $V_{c_1,c_3} = 5000$ and $V_{c_2,c_3} = 1000$, with $V_{min} = 1000$. The expected number of ballot checks required by MACRO to audit this election is 145. In the final round election, our winner is $c_1$ and loser $c_2$, with $V_{c_1,c_2} = 3000$.  The expected number of ballot checks required by MACRO to audit this election is 48.3. The overall ASN for the comparison audit, given simultaneous elimination of candidates $c_4$ and $c_5$, is 145. This is less than that of the ballot-polling variant at 1402.
\end{example} 

\subsection{Winner only auditing}\label{sec:WinnerOnly}

The above two methods consider auditing the entire IRV process to ensure that we
are confident on all its outcomes -- i.e., that the correct candidate was eliminated in each round.  This is too strong since even if earlier
eliminations happened in a different order it may not have any effect on the
eventual winner.

\begin{example} \label{ex:only}
  Consider an election with ballots $[c_1,c_2,c_3]: 10000$, $[c_2,c_1,c_3]:6000$ and
  $[c_3,c_1,c_2]: 5999$. No simultaneous elimination is possible, and auditing
  that $c_3$ is eliminated before $c_2$ will certainly require a full
  recount.  But even if $c_2$ were eliminated first it would not change the
  winner of the election.
\end{example}

\subsubsection{Winner only auditing via ballot-polling}
An alternate approach for auditing IRV
elections is to simply reject the $n-1$ null hypotheses
$\{ \tilde{f}(w) \leq \tilde{\tally}_{\{w,l\}}(l) \}$
where $w$ is the declared winner of the IRV election, and $l \in \cand \setminus \{w\}$ refers to each of the non-winning candidates. 
This hypothesis states that loser $l$ gets more votes than winner $w$ where $l$ is given
the maximal possible votes it could ever achieve before $w$ is eliminated, and
$w$ gets only its first round primary vote (the minimal possible votes it could
ever hold).  When we reject this hypothesis we are confident that there
could not be any elimination order where $w$ is eliminated before $l$.
If all these hypotheses are rejected then we are assured that $w$ is the
winner of the election, independent of the order in which all other candidates are eliminated.

\begin{example} 
  Consider the election of Example~\ref{ex:only}. We must reject the
  hypotheses that $\{ \tilde{f}(c_1) \leq \tilde{\tally}_{\{c_1,c_2\}}(c_2) \}$
  ($c_1$ is eliminated before $c_2$) and
  $\{ \tilde{f}(c_1) \leq \tilde{\tally}_{\{c_1,c_2\}}(c_3) \}$
  ($c_1$ is eliminated before $c_3$).
  The primary vote for $c_1$ is
  10000, while the maximum tally that $c_2$ can achieve before $c_1$ is
  eliminated is 6000. The maximum tally that $c_3$ can
  achieve before $c_1$ is eliminated is 5999. Auditing to reject these
  hypotheses is not difficult. The ASNs are 98.4 and 98.3 ballots. 
  Note that if the $[c_2,c_1,c_3]$ ballots were changed to 
  $[c_2,c_3,c_1]$, the maximum tally that $c_3$ can achieve is 12000,
  and the hypothesis that ($c_1$ is eliminated before $c_3$) could not be
  rejected. In this case just changing a single vote could result in
  $c_3$ winning the election, so this election will need a full
   recount.
\end{example}

\subsubsection{Winner only auditing via a ballot-level comparison RLA}
The ballot-level comparison RLA version of the winner-only audit proceeds in a similar fashion to its ballot-polling counterpart. Given a election with winner $w$ and losers $\cand \setminus \{w\}$, the ballot-polling audit executes a BRAVO audit for each winner-loser $(w,l)$ pair, where $l \in \cand \setminus \{w\}$. In each of these audits, $w$ is awarded only their primary vote $\tilde{f}(w)$, while $l$ is awarded all votes in which they appear before $w$, or where they appear and $w$ does not $\tilde{t}_{\{w,l\}}(l)$. This audit is designed to disprove the null hypothesis that $\tilde{f}(w) \leq \tilde{t}_{\{w,l\}}(l)$. 

We apply the MACRO algorithm of Figure \ref{fig:macro}, in place of BRAVO, for each winner-loser pair $(w,l)$, with  $\winners = \{w\}$ and $\losers = \{l\}$.

\begin{example} 
  Consider the election of Example~\ref{ex:only}. For winner-loser pair $(c_1,c_3)$, we apply MACRO to an election with winner $c_1$, and loser $c_3$, where $c_1$ has a tally of 10000 and $c_3$ a tally of 5999. Even though $c_1$ appears before $c_3$ in the [$c_2$, $c_1$, $c_3$] ballots, we only award $c_1$ with its first preference votes in a winner-only audit. If the positions of $c_1$ and $c_3$ were swapped in these ballots, these ballots would be treated as votes for $c_3$. In this election, $V_{min} = 4001$ and we expect to check 36.2 ballots.  For winner-loser pair $(c_1, c_2)$, we apply MACRO to an election with winner $c_1$, and loser $c_2$, where $c_1$ has a tally of 10000 votes and $c_2$ a tally of 6000. The ASN for this election is also 36.2 ballots.  
\end{example}

\section{Finding efficient ballot-polling or comparison RLAs for IRV} \label{sec:GeneralRLABPCP}

In each of the ballot-polling and ballot-level comparison RLAs for IRV described above, we apply an existing risk limiting audit (BRAVO, as per Figure~\ref{fig:irvbravo}, or MACRO, as per Figure~\ref{fig:macro}) to confirm a collection of assertions with a given level of statistical confidence. In the case of a winner-only audit, for example, we are seeking to confirm that the reported winner $w$ could not have been eliminated before any one of the reported losers $l \in \cand \setminus \{w\}$. This results in $|\cand \setminus \{w\}|$ assertions to be confirmed, one for each winner-loser pair. 

For each assertion that we seek to confirm, we can estimate the number of ballots that must be checked via an application of BRAVO or MACRO, assuming no errors are found.  We present a general algorithm for choosing the set of assertions $\mathcal{A}$ that can be checked most efficiently  to confirm that the reported winner $c_w$ was correct.  The algorithm, denoted RAIRE, achieves this by finding the easiest way to show that all election outcomes in which a candidate other than $c_w$ won, did not arise, with a given level of statistical confidence, for a given method of auditing each assertion. The algorithm can be applied to generate either a ballot-polling or a ballot-level comparison RLA for an IRV election. 

Note that our \emph{risk-limit} follows directly from BRAVO and MACRO: if the election outcome is wrong, then one of the assertions in $\mathcal{A}$ must be false---a BRAVO or MACRO audit with risk limit $\alpha$ will detect this with probability at least $1-\alpha$, and we then manually recount the whole election. 

RAIRE, outlined in Figure~\ref{alg:audit-irv}, explores the tree of alternate elimination sequences, ending in a candidate $c' \neq c_w$. Each node is a partial (or complete) elimination sequence. For each node $\pi$, we consider the set of assertions that (i) can be proven with an application of BRAVO or MACRO and (ii) any one of which disproves the outcome that $\pi$ represents. We label each node $\pi$ with the assertion $a$ from this set that requires the least number of anticipated ballot samples (ASN) to prove, denoted $ASN(a)$. We use the notation $a(\pi)$ and $ASN(a(\pi))$ to represent the assertion assigned to $\pi$ and the ASN for this assertion, respectively. RAIRE finds a set of assertions to prove, denoted $\mathcal{A}$, that: validates the correctness of a given election outcome, with risk limit $\alpha$; and for which the largest ASN of these assertions is minimised. When performing a ballot-polling audit, we compute this ASN via Equation~\ref{eqn:ASNBP}, and Equation~\ref{eqn:ASNCP} when generating a comparison RLA.

Consider a partial elimination sequence $\pi = [c, \ldots, w]$ of at least
two candidates, leading to an alternate winner $w$.
This sequence represents the suffix of a complete order -- an outcome in which the candidates in $\cand\setminus\pi$ have been previously eliminated, in some order. We define a function \textsf{FindBestAudit}($\pi$, $\cand$, $\mathcal{B}$, $\alpha$[, $\gamma$]) that finds the easiest to prove assertion $a$, with the smallest ASN, which disproves the outcome $\pi$ given risk limit $\alpha$. The parameter $\gamma$ is only used when generating a ballot-level comparison RLA. 

For the outcome $\pi = [c|\ldots]$, \textsf{FindBestAudit} considers the following assertions:  
\begin{description}
\item[WO($c$,$c'$):] Assertion that $c$ beats $c' \in \pi$, where $c' \neq c$ appears later in $\pi$, in a winner
  only audit of the form described in Section \ref{sec:WinnerOnly}, with
  winner $c$ and loser $c'$. This invalidates the sequence as $c$ cannot
  be eliminated before $c'$; \\
\item[WO($c''$,$c$):] Assertion that $c'' \in \cand \setminus \pi$ beats $c$ in a winner-only audit with
  winner $c''$ and loser $c$. Candidate $c''$ does not appear in $\pi$, and consequently is assumed eliminated in some prior round. Proving this assertion invalidates the sequence as it shows that $c''$ 
  cannot be eliminated before $c$;\\
\item[IRV($c$,$c'$,$\{ c'' ~|~ c'' \in \pi$\}):] Assertion that  $c$ beats some $c' \neq c \in \pi$ in a BRAVO (or MACRO) audit with winner $c$
  and loser $c'$, under the assumption that the only candidates remaining
  are those in $\pi$ (i.e. the set $\{c'' ~|~ c'' \in \pi\}$).
  All other candidates have eliminated with their votes distributed to later preferences. Proving this assertion
  invalidates the sequence as shows that $c$ cannot be eliminated at this
  stage in an IRV election.
\end{description}
We assume that if no assertion exists with an ASN less than $|\mathcal{B}|$ the
function returns a dummy $\textbf{INF}$ assertion with $ASN(\textbf{INF}) = +\infty$.

\begin{figure}[!t]
\begin{singlespace}
  \begin{ctabbing}
    xxx \= xx \= xx \= xx \= xx \= xx \= xx \= xx \= xx \= \kill
    RAIRE($\cand$, $\mathcal{B}$, $c_w$, $\alpha$[, $\gamma$]) \\
1    \> $\mathcal{A} \leftarrow \emptyset$ \\[2pt]
2	\> $F \leftarrow \emptyset$ $\triangleright$ $F$ is a set sequences
to expand (the frontier) \\
3	\> $LB \leftarrow 0$ \\
	\> $\triangleright$ Populate $F$ with single-candidate sequences \\
4	\> \textbf{for each}($c \in \cand\setminus\{c_w\}$): \\
5	\> \> $\pi \leftarrow [c]$ \\
6	\> \> $a$ $\leftarrow$ \textsf{FindBestAudit}($\pi$, $\cand$, $\mathcal{B}$, $\alpha$[, $\gamma$]) \\
7       \> \> $asr[\pi] \leftarrow a$ $\triangleright$ Record best assertion
for $\pi$ \\
8       \> \> $ba[\pi] \leftarrow \pi$ $\triangleright$ Record best ancestor
        sequence for $\pi$ \\
9	\>\> $F \leftarrow F \cup \{ \pi \}$ \\
	\> $\triangleright$ Repeatedly expand the sequence with largest ASN in $F$ \\
10	\> \textbf{while}($|F| > 0$): \\
11      \>\> $\pi \leftarrow \textsf{argmax} \{ ASN(asr[\pi]) ~|~ \pi \in F \}$
        \\
12      \>\> $F \leftarrow F \setminus \{ \pi \}$ \\
13	\>\> \textbf{if}($ASN(asr[ba[\pi]]) \leq LB$):\\
14      \>\>\> $\mathcal{A} \leftarrow \mathcal{A} \cup \{ asr[ba[\pi]] \}$ \\
15	\>\>\> $F$ $\leftarrow$ $F \setminus \{ \pi' \in F ~|~ ba[\pi]$ is a suffix of $\pi' \}$ \\
16	\>\>\> \textbf{continue} \\
17	\>\> \textbf{for each}($c \in \mathcal{C}\setminus\pi$): \\
18	\>\>\> $\pi' \leftarrow [c]\plusplus\pi$ \\
19	\>\>\> $a$ $\leftarrow$ \textsf{FindBestAudit}($\pi'$, $\cand$,
$\mathcal{B}$, $\alpha$[, $\gamma$]) \\
20      \>\>\> $asr[\pi'] \leftarrow a$ \\
21	\>\>\> $ba[\pi']$ $\leftarrow$ \textbf{if} $ASN(a) < ASN(asr[ba[\pi]])$
\textbf{then} $\pi'$ \textbf{else} $ba[\pi]$ \\
22	\>\>\> \textbf{if}($|\pi'| = |\cand|$): \\
23	\>\>\>\> \textbf{if}($ASN(asr[ba[\pi']]) = \infty$): \\
24	\>\>\>\>\> \textbf{terminate} algorithm, full recount necessary\\[2pt]
25	\>\>\>\> \textbf{else}: \\
26      \>\>\>\>\> $\mathcal{A} \leftarrow \mathcal{A} \cup \{ asr[ba[\pi']] \}$ \\
27	\>\>\>\>\> $LB$ $\leftarrow$ $max$($LB$, $ASN(asr[ba[\pi']])$) \\
28	\>\>\>\>\> $F$ $\leftarrow$ $F \setminus \{ \pi' \in F ~|~ ba[\pi]$ is
a suffix of $\pi' \}$ \\
29      \>\>\>\>\> \textbf{continue} \\
30	\>\>\> \textbf{else}: \\
31	\>\>\>\> $F \leftarrow F \cup \{ \pi' \}$ \\
32	\> \textbf{return} $\mathcal{A}$ with maximum ASN equal to $LB$
  \end{ctabbing}
\end{singlespace}
\caption{The RAIRE algorithm for finding a least cost collection of assertion s to audit, with parallel applications of BRAVO or MACRO, defining a risk-limiting audit of an IRV election with candidates $\cand$, ballots $\mathcal{B}$, and winner $c_w$, with a given risk limit $\alpha$, and inflator parameter $\gamma$ (if using MACRO). }
\label{alg:audit-irv}
\end{figure}

For an election with candidates $\cand$ and winner $c_w$, RAIRE
starts by adding $|\cand| - 1$ partial elimination orders to an initially
empty priority queue $F$, one for each alternate winner $c \neq c_w$ (Steps
4 to 9).  The set $\mathcal{A}$ is initially empty.
For orders $\pi$ containing a single candidate $c$,
\textsf{FindBestAudit} considers the assertions \textbf{WO}($c''$,$c$),
which means that candidate $c'' \neq c$
beats $c$ in a winner-only audit of the form described in Section
\ref{sec:WinnerOnly}, with winner $c''$ and loser $c$, for each $c'' \in
\cand \setminus \{c\}$.
The assertion $a$ with the smallest $ASN(a)$ is
recorded in $asr[\pi]$ (Step 7). For each partial sequence $\pi$ added to $F$, we are finding the easiest to prove assertion $a$ that invalidates the outcome. 

For each partial elimination order $\pi$ in $F$, we also keep record of a `best ancestor' $ba[\pi]$ (Steps 8 and 21). The best ancestor is the subsequence of $\pi$ that can be invalidated with the least auditing effort. For example, consider a partial elimination order $\pi$ $=$ [$c_1,$ $c_3,$ $c_2$]. This corresponds to an election in which $c_1$, $c_3$, and $c_2$ are the last three remaining candidates, after all others have been eliminated, with $c_2$ winning the election.
The ancestors of $\pi$ are the sequences [$c_1,$ $c_3,$ $c_2$], [$c_3,$ $c_2$] and [$c_2$]. Note that we include the sequence itself in this set. The best ancestor is the (sub)sequence $\pi'$ with the smallest $asr[\pi']$. This tells us that proving assertion $asr[\pi']$ is currently the best known way to disprove all elimination orders that end in $\pi$.  
 The purpose of keeping track of these `best ancestors' will become apparent in the following paragraphs. 

The algorithm works by exploring a tree of possible alternate elimination orders. For each elimination order that it visits, it finds the least cost assertion that, if proven, invalidates the outcome. 
As the algorithm progresses, we maintain a lower bound (initialised to 0 in Step 3) on the total auditing effort required in the most efficient audit for the given election. This corresponds to the ASN of the hardest to prove assertion sitting in $\mathcal{A}$.  

We repeatedly find and remove a partial sequence $\pi$ in $F$ for
expansion (Steps 11 and 12). This is the sequence with the (equal) highest ASN -- i.e., the sequence $\pi$ with the highest associated $asr[\pi]$.
When expanding a sequence $\pi$, we first check whether its current best ancestor has an ASN lower than the current
lower bound $LB$ (Steps 13 to 16). If so, 
we simply add the corresponding assertion ($asr[ba[\pi]]$) to $\mathcal{A}$
and remove any sequences in $F$ which are subsumed by this ancestor (have it
as a suffix). We know that the best audit we can find has an ASN of at least $LB$. Adding $asr[ba[\pi]]$ to our audit will disprove all outcomes ending in $ba[\pi]$ while not increasing its ASN.

Otherwise, 
we create a new elimination sequence $\pi'$ with $c$
appended to the start of $\pi$ ($[c]\plusplus\pi$) for each $c \in \cand
\setminus \pi$ (Steps 17 to 31). 
For a new sequence $\pi'$, \textsf{FindBestAudit} finds the assertion $a$ requiring the least auditing effort to prove in order to invalidate $\pi'$.
We record (Step 20) this as the assertion for $asr[\pi'] = h$.
We calculate (Step 21) the best ancestor of $\pi'$ by comparing
the ASN for its assertion with that of its best ancestor.
If the sequence $\pi'$ is complete, then we known one of its ancestors
(including itself) must be audited.  If the best of these is infinite, we
terminate, a full recount is necessary.
Otherwise we add the assertion of its best ancestor to $\mathcal{A}$ and remove
all sequences in $F$ which are subsumed by this ancestor. We update $LB$ to the ASN of the hardest to prove assertion already part of our audit (Step 27). 
If the sequence is not complete, we add it to our frontier $F$.

\begin{example} \label{ex:genalg}
  Consider an election with ballots 
  $[c_1,c_2,c_3]:5000$, 
  $[c_1,c_3,c_2]:5000$	  
  $[c_2,c_3,c_1]:5000$,
  $[c_2,c_1,c_3]:1500$,
  $[c_3,c_2,c_1]:5000$,
	$[c_3,c_1,c_2]:500$, and $[c_4,c_1]:5000$, and candidates $c_1$ to $c_4$. The initial tallies are: $c_1$: 10000; $c_2$: 6500; $c_3$: 5500; $c_4$: 5000. Candidates $c_4$, $c_3$, and $c_2$ are eliminated, in that order, with winner $c_1$. 

In a ballot-polling or comparison winner-only audit ($\alpha=0.05$), we cannot show that $c_1$ beats $c_3$, or that $c_1$ beats $c_2$, as $c_1$'s first preference tally (of 10000 votes) is less than the total number of ballots that we could attribute to $c_2$ and $c_3$ (11500 and 10500, respectively). Simultaneous elimination is not applicable, as no sequences of candidates can be eliminated in a group. In an audit of the whole elimination order (as per Section \ref{sec:particularElim}), the loss of $c_4$ to $c_1$, $c_2$, and $c_3$ is the most challenging to audit. 
The ASN for the ballot-polling version of this audit, assuming $\alpha = 0.05$, is 25\% of all ballots (6750 ballots). The ballot-level comparison version of this audit, assuming $\gamma = 1.1$, is 1.3\% of all ballots (351 ballots). 

	RAIRE finds a set of assertions that can be proven using a ballot-polling audit with a maximum ASN of 1\% (or 270 ballots, with $\alpha=0.05$), and that consequently rule out all elimination sequences that end in a candidate other than $c_1$. This audit tests the assertions: 
 $c_1$ beats $c_2$ if $c_3$ and $c_4$ have been eliminated (ASN of 1\%); 
$c_1$ beats $c_3$ if $c_2$ and $c_4$ have been eliminated (ASN 0.5\%); 
$c_1$ beats $c_4$ in a winner-only audit (ASN 0.4\%); and 
that $c_1$ beats $c_3$ if $c_4$ has been eliminated (ASN 0.1\%). 

If we instead use RAIRE to build a comparison audit, we find an audit with a maximum ASN of 0.17\%. The assertions in this audit are: 
$c_1$ beats $c_2$ if $c_3$ and $c_4$ have been eliminated (ASN of 0.17\%);
$c_1$ beats $c_3$ if $c_4$ has been eliminated (ASN of 0.07\%); 
$c_1$ beats $c_3$ if both $c_2$ and $c_4$ has been eliminated (ASN of 0.11\%); 
$c_1$ beats $c_4$ in a winner-only audit (ASN of 0.13\%); and 
$c_2$ and $c_3$ beat $c_4$ with all remaining candidates eliminated (ASN of 0.04\%). 
\end{example}

\section{Computational Results}\label{sec:CompResults}

We have simulated the ballot-polling and ballot-level comparison RLAs described in Section \ref{sec:OurAudits} on 21 US IRV elections held between 2007 and 2014, and on the IRV elections held across 93 electorates in the 2015 state election in New South Wales (NSW), Australia. For each election, we have simulated each audit with varying risk limits ($\alpha = 1\%$ and $\alpha = 5\%$), and $\gamma = 1.1$.\footnote{We explore the influence of the $\gamma$ parameter in subsequent experiments.} We record, for each simulated audit, the number of ballots that were sampled during the audit (expressed as a percentage of ballots cast). An audit that needs to sample fewer ballots before confirming the correctness of the reported outcome, to the given degree of statistical confidence, is a more efficient audit. As each audit involves ballots being drawn at random, we simulate each audit 10 times and compute the average number of ballots checked across those 10 simulations.

All experiments have been conducted on a machine with an Intel Xeon Platinum 8176 chip (2.1GHz), and 1TB of RAM. The aim of these experiments is to demonstrate that RLAs for IRV elections are feasible -- only a small fraction of ballots need be sampled in elections that are not extremely close -- and compare the efficiency of different types of IRV RLA. 

Table \ref{tab:BPvsCP_EO} compares the number of ballot checks required by ballot-polling and  comparison audits of the form described in Section~\ref{sec:particularElim} (auditing the entire elimination order) across our suite of election instances. The number of required samples is reported alongside the ASN for each audit (computed as per Equation~\ref{eqn:ASNBP} for each ballot-polling audit, and Equation~\ref{eqn:ASNCP} for each comparison audit), and the margin of victory (MOV) for each election (computed using the algorithm of \cite{blom16}).
Tables~\ref{tab:BPvsCP_SE} and~\ref{tab:BPvsCP_WO} similarly compare the number of ballot samples required by simultaneous elimination and winner-only ballot-polling and comparison audits, described in Sections~\ref{sec:Batch} and~\ref{sec:WinnerOnly}. In this experiment, no errors or discrepancies have been injected into the set of reported ballots in each election instance ($\ballots \equiv \tilde{\ballots}$).

\afterpage{
\begin{landscape}
\begin{table}[!t]
\footnotesize
\begin{tabular}{|ll|c|c|c||cc|cc||cc|cc|}
\multicolumn{13}{r}{\textbf{Auditing the Entire Elimination Order via a Ballot Polling (BP) and Comparison (CP) Audit}} \\
\hline
      &   & &     &     & \multicolumn{4}{c||}{$\alpha$ $=$ 1\% } & \multicolumn{4}{c|}{$\alpha$ $=$ 5\% }  \\
\cline{6-13}
      &   & &     &     & \multicolumn{2}{c|}{ BP } & \multicolumn{2}{c||}{ CP ($\gamma = 1.1$) } & \multicolumn{2}{c|}{ BP } & \multicolumn{2}{c|}{ CP ($\gamma = 1.1$) }  \\
\cline{6-13}
\multicolumn{2}{|c|}{\textbf{Election}} & $|\cand|$ & $|\mathcal{B}|$ & \textbf{MOV} & Polls \%  & ASN \%  & Polls \% & ASN \%  & Polls \%  & ASN \%  & Polls \% & ASN \%  \\
\hline 
1 & Berkeley 2010 D7 CC    & 4 & 4,682    & 364 (7\%)      & 6.7 & 7.2                   & \bf 1.6 & \bf 1.7 &  3.9 & 4.7  & \bf 1.1& \bf 1.1\\
2 & Berkeley 2010 D8 CC    & 4 & 5,333    & 878 (16\%)     & $\infty$ & $\infty$         &$\infty$ & $\infty$ & $\infty$ & $\infty$ & \bf 94.1& \bf 94.2 \\
3 & Oakland 2010 D6 CC     & 4 & 14,040   & 2,603 (19\%)   & 4.0 & 4.4                   & \bf 1 & \bf 1 & 3 & 2.9  & \bf 0.7 & \bf 0.7\\
4 & Pierce 2008 CC         & 4 & 43,661   & 2,007 (5\%)    & 3.1 & 2.2                   & \bf 0.2& \bf 0.3 & 1.8 & 1.4 & \bf 0.2 & \bf 0.2 \\
5 & Pierce 2008 CAD        & 4 & 159,987  & 8,396 (5\%)    & 0.3 & 0.5                   & \bf 0.1 & \bf 0.1 & 0.2 & 0.3 & \bf 0.1& \bf 0.1 \\
6 & Aspen 2009 Mayor       & 5 & 2,544    & 89 (4\%)       & 62.4 & 71.8                 & \bf 9 &  \bf 9 & 52.7 & 46.9 & \bf 5.9 & \bf 5.9 \\ 
7 & Berkeley 2010 D1 CC    & 5 & 6,426    & 1,174 (18\%)   & \bf 2.4 & \bf 1.7                   & 5.8 & 5.9  & \bf 1.6 & \bf 1.1 &3.8 & 3.8\\
8 & Berkeley 2010 D4 CC    & 5 & 5,708    & 517 (9\%)      & 7.5 & 7                     & \bf 6.3 & \bf 6.3  & 6 & 4.7  & \bf 4.1& \bf 4.1\\
9 & Oakland 2012 D5 CC     & 5 & 13,482   & 486 (4\%)      & 11.2 & 10.3                 & \bf 1.5 & \bf 1.5  & 7.3 & 6.7 & \bf 1 & \bf 1\\ 
10 & Pierce 2008 CE         & 5 & 312,771  & 2,027 (1\%)    & 11.6 & 15.1                 & \bf 0.2 & \bf 0.2 & 7.6 & 9.8 & \bf 0.2 & \bf 0.2\\
11 & San Leandro 2012 D4 CC&5& 28,703& 2,332 (8\%)& 9.3 & 9.7               & \bf 0.9 & \bf 1  & 6.3  & 6.3 & \bf 0.6 & \bf 0.6\\
12 & Oakland 2012 D3 CC     & 7 & 26,761   & 386 (1\%)      & $\infty$ & $\infty$         & \bf 22.5& \bf 22.5& $\infty$ & $\infty$ & \bf 14.6& \bf 14.7\\
13 & Pierce 2008 CAS        & 7 & 312,771  & 1,111 (0.4\%)  & $\infty$ & $\infty$         & \bf 3.7 & \bf 3.7 & $\infty$ & $\infty$ & \bf 2.4& \bf 2.4\\
14 & San Leandro 2010 Mayor & 7 & 23,494 & 116 (0.5\%)& $\infty$ &$\infty$                &  \bf 18.4  & \bf 18.4   & 92.9 & $\infty$ & \bf 12 & \bf 12 \\
15 & Berkeley 2012 Mayor    & 8 & 57,492   & 8,522 (15\%)   & 94.6 & $\infty$             &  \bf 5.9  & \bf 5.9 & 77 & $\infty$  & \bf 3.8 & \bf 3.8\\
16 & Oakland 2010 D4 CC     & 8 & 23,884   & 2,329 (10\%)   & $\infty$  &$\infty$         &  \bf 7.7  & \bf 7.7 & 76.4  & $\infty$  & \bf 5.4& \bf 5.4\\ 
17 & Aspen 2009 CC          & 11& 2,544    & 35 (1\%)       & $\infty$ &$\infty$          & $\infty$ & $\infty$  & $\infty$ &$\infty$ & $\infty$& $\infty$\\
18 & Oakland 2010 Mayor     & 11& 122,268  & 1,013 (1\%)    & $\infty$ &$\infty$          & \bf 12.2 & \bf 12.2 & $\infty$ &$\infty$  &\bf 7.9 & \bf 7.9\\
19 & Oakland 2014 Mayor     & 11& 101,431  & 10,201 (10\%)  & $\infty$ &$\infty$          & $\infty$ & $\infty$ & $\infty$ &$\infty$ & $\infty$& $\infty$ \\
20 & San Francisco 2007 Mayor& 18& 149,465 & 50,837 (34\%)&$\infty$&$\infty$  & $\infty$ & $\infty$  & $\infty$ &$\infty$ & $\infty$& $\infty$\\
21 & Minneapolis 2013 Mayor  & 36& 79,415  & 6,949 (9\%) & $\infty$&$\infty$& $\infty$ & $\infty$ & $\infty$ &$\infty$ & $\infty$ & $\infty$ \\
\hline
22 & Balmain NSW 2015                 & 7 & 46,952 & 1,731 (3.7\%) &  $\infty$ &$\infty$  & \bf 34.9 & \bf 34.9 & $\infty$ &$\infty$ & \bf 22.7 & \bf 22.7 \\
23 & Campbelltown NSW 2015  & 5 & 45,124 & 3,096 (6.9\%) & 13.6 &12.2       &\bf 2.2  & \bf 2.2 & 8.4 &8  &\bf  1.4&\bf  1.4\\
24 & Gosford NSW 2015                 & 6 & 48,259 & 102 (0.2\%)   &  $\infty$ &$\infty$  & \bf 6.6  &\bf  6.6  & $\infty$ &$\infty$ &\bf 4.3 &\bf 4.3\\
25 & Lake Macquarie NSW 2015 & 7 & 47,698 & 4,253 (8.9\%) & 27.7 & 22.8     &\bf 3.5 &\bf 3.5 & 14.5 & 15 & \bf 2.3& \bf 2.3 \\
26 & Sydney NSW 2015                 & 8 & 42,747 & 2,864 (6.7\%) & $\infty$ & $\infty$   & \bf 59.6 & \bf 59.6 & $\infty$ & $\infty$ & \bf 38.8 &\bf  38.8  \\
\hline
\end{tabular}
\caption{Average \# of ballots sampled (as a percentage of ballots cast) over 10 simulated ballot-polling (BP) and ballot-level comparison audits (CP) of 26 IRV elections using the EO (auditing the entire elimination order) method. Parameter $\alpha$ ranges between 1\% and 5\%, and $\gamma = 1.1$. Also reported is each election's margin of victory (MOV). The notation $\infty$ indicates a percentage of ballots (or ASN) greater than 100\%. CC, CE, CAD, and CAS denote City Council, County Executive, County Auditor, and County Assessor. The most efficient audit (for $\alpha=$1\% and 5\%) is highlighted in bold.}
\label{tab:BPvsCP_EO}
\end{table}
\end{landscape}}

\afterpage{
\begin{table}[!t]
\footnotesize
\centering
\begin{tabular}{|l|l||cc|cc||cc|cc|}
\multicolumn{10}{r}{\textbf{Auditing with Simultaneous Elimination via a BP/CP Audit}} \\
\hline
	&  & \multicolumn{4}{c||}{$\alpha$ $=$ 1\% } & \multicolumn{4}{c|}{$\alpha$ $=$ 5\% }  \\
\cline{3-10}
	$\#$   & $|\mathcal{B}|$  &  \multicolumn{2}{c|}{ BP (\%)} & \multicolumn{2}{c||}{ CP ($\gamma = 1.1$, \%) } & \multicolumn{2}{c|}{ BP (\%) } & \multicolumn{2}{c|}{ CP ($\gamma = 1.1$, \%) }  \\
\cline{3-10}
	&  & Polls & ASN  & Polls & ASN   & Polls   & ASN   & Polls  & ASN  \\
\hline 
	1 & 4,682 & 7.5 & 7.2          &\bf 1.3  & \bf 1.4  & 4  & 4.7&\bf 0.9 &\bf 0.9 \\
	2 & 5,333 & 2.9 & 4.2          & \bf 1 & \bf 1  & 2  & 2.8& \bf 0.6 &\bf  0.7 \\
	3 & 14,040 & 0.7 & 0.9          & \bf 0.3  & \bf 0.3  & 0.5&0.6 & \bf 0.2 &\bf  0.2 \\
	4 & 43,661 & 3.1 & 2.2          & \bf 0.3 &\bf  0.3  & 1.8&1.4 &\bf 0.2&\bf 0.2  \\
	5 & 159,987 & 0.3 & 0.5          &\bf 0.1  & \bf 0.1  &0.2&0.3 &\bf  0.04&\bf 0.04 \\
	6 & 2,544 & 62.4& 71.8         &\bf 5.7 &\bf  5.7 &54.8&46.9&\bf 3.7&\bf 3.7 \\ 
	7 & 6,426 & 2.4 & 1.7          &\bf 0.5 &\bf 0.6  &1.6 &1.1 &\bf 0.4&\bf 0.4\\
	8 & 5,708 & 28.7& 40.7         & \bf 3 &\bf 3.1  &17.8&26.6&\bf  2 &\bf 2 \\
	9 & 13,482 & 15.1& 10.3         &\bf 1.1 &\bf 1.1  &11.8&6.7 &\bf 0.7&\bf 0.7 \\ 
	10 &312,771  & 11.6& 15.1         &\bf 0.2 &\bf 0.2  &7.6 &9.8 &\bf 0.2&\bf 0.2 \\
	11 &28,703  & 9.3 & 9.7          & \bf 0.9 &\bf 1  &6.3 &6.3 &\bf 0.6&\bf 0.6 \\
	12 &26,761  & $\infty$ & $\infty$& \bf 22.5 &\bf 22.5  &$\infty$&$\infty$  &\bf 14.7&\bf 14.7 \\
	13 &312,771 & $\infty$ & $\infty$& \bf 3.7 &\bf  3.7 &$\infty$&$\infty$  &\bf 2.4 &\bf 2.4 \\
	14 & 23,494 & $\infty$ & $\infty$&\bf 4.4 &\bf 4.4  &92.9&$\infty$ &\bf 2.8&\bf 2.8 \\
	15& 57,492 & 2.3 & 2.6          &\bf 0.2 &\bf 0.2  &1.6 &1.7  &\bf 0.2 &\bf 0.2 \\
	16& 23,884 & $\infty$ & $\infty$&\bf 7.7 &\bf 7.7  &$\infty$&$\infty$  &\bf 5&\bf 5  \\ 
	17& 2,544 & $\infty$ & $\infty$& $\infty$ & $\infty$  &$\infty$&$\infty$  & \bf 82.4& \bf 82.4 \\
	18& 122,268 & 21.5 & 23.8        & \bf 0.6 & \bf 0.6  &15 & 15.5 & \bf 0.4& \bf 0.4 \\
	19&101,431 & $\infty$ &$\infty$ & $\infty$ & $\infty$  &$\infty$&$\infty$  & $\infty$& $\infty$  \\
	20&149,465 & 0.03 & 0.03        & \bf 0.01 & \bf 0.01  &0.02 & 0.02  &\bf  0.01&\bf  0.01 \\
	21&79.415 & $\infty$ & $\infty$& $\infty$ & $\infty$  &$\infty$ & $\infty$  & $\infty$& $\infty$  \\
\hline
	22& 46,952 & 83.8 & $\infty$    & \bf 2.2 &\bf  2.2  & 65.4 & 82 &\bf 1.4&\bf  1.4 \\
	23& 45,124& $\infty$ &$\infty$ &\bf 3.7 &\bf 3.7  & $\infty$ & $\infty$ &\bf 2.4 &\bf 2.4 \\
	24& 48,259& $\infty$ &$\infty$ & \bf 5 & \bf 5  & $\infty$ & $\infty$ &\bf 3.2 &\bf 3.2 \\
	25& 47,698& 6.9 & 7.8          & \bf 0.6 &\bf 0.6  & 3.2 & 5.1 &\bf 0.4 &\bf 0.4 \\
	26& 42,747 & 3.3 & 4.6          &\bf 0.4 &\bf 0.4  & 2.2 & 3 &\bf 0.3&\bf 0.3  \\
\hline
\end{tabular}
\caption{Average \# of ballots sampled (as a percentage of ballots cast) over 10 simulated ballot-polling (BP) and comparison audits (CP) of 26 IRV elections with simultaneous elimination (SE), $\alpha$ $\in \{1\%,5\%\}$, and $\gamma = 1.1$. A $\infty$ indicates a percentage of ballots (or ASN) greater than 100\%. The name, candidates, and MOV of each election are shown in Table~\ref{tab:BPvsCP_EO}.  The most efficient audit (for $\alpha=$1\% and 5\%) is highlighted in bold.}
\label{tab:BPvsCP_SE}
\end{table}
}

\afterpage{
\begin{table}[!t]
\footnotesize
\centering
\begin{tabular}{|l|l||cc|cc||cc|cc|}
\multicolumn{10}{r}{\textbf{Winner-Only Auditing via a BP/CP Audit}} \\
\hline
	&   &  \multicolumn{4}{c||}{$\alpha$ $=$ 1\% } & \multicolumn{4}{c|}{$\alpha$ $=$ 5\% }  \\
\cline{3-10}
	$\#$  & $|\mathcal{B}|$    &  \multicolumn{2}{c|}{ BP (\%)} & \multicolumn{2}{c||}{ CP ($\gamma = 1.1$, \%) } & \multicolumn{2}{c|}{ BP (\%)} & \multicolumn{2}{c|}{ CP ($\gamma = 1.1$, \%) }  \\
\cline{3-10}
	& & Polls  & ASN   & Polls  & ASN   & Polls   & ASN   & Polls  & ASN  \\
\hline 
	1 & 4,682 & 8.7& 22.4          & \bf 2.5 &\bf 2.6  & 4.9  & 14.7&\bf 1.7&\bf 1.7\\
	2 & 5,333 &1.3& 1.8           & \bf 0.7 &\bf 0.7  & 0.8  & 1.2 &\bf 0.4&\bf  0.5  \\
	3 & 14,040 &0.4& 0.5           &\bf 0.2 &\bf 0.2  & 0.3  & 0.3&\bf 0.1 &\bf 0.1  \\
	4 & 43,661 &3.2& 4.1           &\bf 0.3 &\bf 0.3  & 1.8  & 2.7&\bf 0.2&\bf 0.2   \\
	5 & 159,987 & 0.5& 1.2           &\bf 0.1 &\bf 0.1  & 0.3  & 0.8&\bf 0.1&\bf 0.1  \\
	6 & 2,544& $\infty$ & $\infty$ & $\infty$ & $\infty$  & $\infty$  & $\infty$& $\infty$ & $\infty$   \\ 
	7 & 6,426 & 1.1& 1.1           &\bf 0.4 &\bf 0.5  & 0.8  & 0.7&\bf 0.3&\bf 0.3 \\
	8 & 5,708 &4.9& 7.3           &\bf 1.3 &\bf 1.4  & 3.8  & 4.8&\bf 0.9&\bf 0.9 \\
	9 & 13,482 &$\infty$& $\infty$ & $\infty$ & $\infty$  & $\infty$  & $\infty$& $\infty$ & $\infty$   \\ 
	10& 312,771 &$\infty$& $\infty$ & $\infty$ & $\infty$  & $\infty$  & $\infty$& $\infty$ & $\infty$    \\
	11& 28,703 &1.1& 4.4           &\bf 0.5 &\bf 0.5  & 0.8 & 2.9 &\bf 0.3 &\bf 0.3 \\
	12 & 26,761 & $\infty$& $\infty$ & $\infty$ & $\infty$  & $\infty$  & $\infty$& $\infty$ & $\infty$    \\
	13 & 312,771 & $\infty$& $\infty$ & $\infty$ & $\infty$  & $\infty$  & $\infty$&$\infty$  & $\infty$  \\
	14 & 23,494 &$\infty$& $\infty$ & $\infty$ & $\infty$  & $\infty$  & $\infty$& $\infty$ & $\infty$  \\
	15 & 57,492 &0.2& 0.2           & \bf 0.1 &\bf  0.1  & 0.1  & 0.2&\bf 0.1 &\bf 0.1\\
	16 & 23,884 &0.9& 3.1           & \bf 0.5 &\bf 0.5  & 0.6  & 2&\bf 0.3&\bf 0.3  \\ 
	17 & 2,544 &$\infty$& $\infty$ & $\infty$ & $\infty$  & $\infty$  & $\infty$& $\infty$& $\infty$  \\
	18& 122,268 &$\infty$& $\infty$ & $\infty$ & $\infty$  & $\infty$  & $\infty$ & $\infty$& $\infty$  \\
	19& 101,431& 0.8 & 19.8         &\bf 0.6 &\bf 0.6  & 0.5  & 12.9&\bf 0.4& \bf 0.4 \\
	20& 149,465 &0.01& 0.01         & 0.01 & 0.01  & 0.01  & 0.01& 0.01& 0.01 \\
	21& 79,415 &0.5 & 3.1          & \bf 0.3 &\bf 0.3  & 0.3  &2.1 &\bf 0.2&\bf 0.2  \\
\hline
	22& 46,952& 5.2& 31.6          &\bf 1 &\bf 1  & 3.7  & 20.6&\bf 0.7&\bf 0.7 \\
	23 &45,124& 1.3& 1.7           &\bf 0.2 &\bf 0.2  & 0.9  & 1.1&\bf  0.1&\bf 0.1  \\
	24 &48,259& $\infty$& $\infty$ & $\infty$ & $\infty$  & $\infty$  &$\infty$ & $\infty$& $\infty$ \\
	25 &47,698& 0.7& 1.6           & \bf 0.2 &\bf  0.2  & 0.5  & 1& \bf 0.1&\bf  0.1 \\
	26 &42,747& 1.6& 6.9           & \bf 0.5 &\bf 0.5  &  1 & 4.5&\bf 0.3&\bf  0.3  \\
\hline
\end{tabular}
\caption{Average \# of ballots sampled (as a percentage of ballots cast) over 10 simulated ballot-polling (BP) and ballot-level comparison audits (CP) of 26 IRV elections using the winner-only (WO) method, $\alpha$ $\in \{1\%,5\%\}$, and $\gamma = 1.1$. A $\infty$ indicates a percentage of ballots (or ASN) greater than 100\%. The name, candidates, and MOV of each election are shown in Table \ref{tab:BPvsCP_EO}. The most efficient audit (for $\alpha=$1\% and 5\%) is highlighted in bold.}
\label{tab:BPvsCP_WO}
\end{table}
}
Tables \ref{tab:BPvsCP_EO} to \ref{tab:BPvsCP_WO} show that performing a winner-only audit can be much easier than auditing the full elimination order (with or without the use of simultaneous elimination), irrespective of whether we are conducting a ballot-polling or comparison audit. This is the case for the 2013 Minneapolis Mayor and 2014 Oakland Mayor elections. In some cases, winner-only audits are more challenging (or not possible) as we seek to show that a candidate $c$ (on just their first preference votes) could have beaten another $c'$ (who is given all votes in which they appear before $c$ or in which they appear, but $c$ does not). Even if $c$ does beat $c'$ in the true outcome of the election, this audit may not be able to prove this (see Pierce 2008 County Executive, Oakland 2012 D5 City Council, and Aspen 2009 Mayor for examples). 

Auditing with simultaneous elimination (grouping several eliminated candidates into a single `super' candidate) can be more efficient than auditing each individual elimination. This is evident in the context of both ballot-polling audits (see Berkeley 2010 D8 City Council, Berkeley 2012 Mayor, Oakland 2010 Mayor, San Francisco 2007 Mayor, and Sydney NSW) and comparison audits (see Balmain NSW 2015, Sydney NSW 2015, Oakland 2010 Mayor, San Leandro 2010 Mayor, and Berkeley 2010 D8 City Council).
 Across the 26 elections in Tables~\ref{tab:BPvsCP_EO} and~\ref{tab:BPvsCP_SE}, conducting a comparison audit with simultaneous elimination was beneficial in 15 instances and detrimental in 2. In the context of ballot-polling audits, simultaneous elimination was beneficial in 8 and detrimental in 5. In some instances, the tally of the super candidate is quite close to that of the next eliminated candidate, resulting in a more challenging audit. This is particularly evident in the ballot-polling audits of Campbelltown NSW and Berkeley 2010 D4 City Council. 

Tables~\ref{tab:BPvsCP_EO} to~\ref{tab:BPvsCP_WO} show that IRV comparison audits are generally more efficient than their ballot-polling counterparts, as they are for FPTP elections.  The Oakland 2012 D3 City Council election is an excellent example. Neither auditing the entire elimination sequence, the sequence with simultaneous elimination, or conducting a winner-only audit, is successful in the ballot-polling context. The ASN is more than the total number of ballots in each case, indicating that a full recount is required. We can conduct a comparison audit, using each of these methods, however, that requires only a fraction of cast ballots to be sampled (23\% or 6155 ballots, 23\%, and 0.1\% or 268 ballots, when auditing the entire elimination order, auditing with simultaneous elimination, and conducting a winner-only audit, respectively). For each simulated audit, increasing the risk limit reduced the average number of required ballot samples, as expected.

Table \ref{tab:AUDITIRV-NOERRORS} reports the average number of ballots examined by the ballot-polling and ballot-level comparison audits generated by RAIRE across the 26 considered elections (with $\alpha = 5$\%). We compare this level of auditing effort against the number of ballot checks required by the best alternate auditing method (auditing the entire elimination order [EO], simultaneous elimination [SE], and winner-only auditing [WO]). Recall that RAIRE finds an appropriate set of assertions to audit (via ballot-polling or a comparison audit) that, if shown to hold with a given degree of statistical confidence, confirms the reported election outcome with that degree of statistical confidence. The algorithm finds the set of assertions requiring the least anticipated number (ASN) of ballot checks to confirm. Table \ref{tab:AUDITIRV-NOERRORS} shows that while the ASN of the RAIRE audits is minimal -- the actual level of auditing effort required by these audits will differ from these estimates, and may be greater than that required by an EO, SE, or WO audit. For ballot polling audits the discrepancy can be large.  In these experiments we have not introduced any errors or discrepancies between the electronic ballot records and the paper ballots. In this setting, the ASN computed for a ballot-level comparison audit accurately represents the actual number of ballot checks or polls made during the audit. 

In all but one of the elections in Table \ref{tab:AUDITIRV-NOERRORS}, RAIRE is able to compute an audit configuration in less than 1 minute. The algorithm requires between 0.003s and 106s to find an audit configuration in the ballot-polling context, and 0.002s to 139s in the comparison audit context. The most time consuming instance is the 2014 Oakland Mayoral election, with RAIRE requiring 106s and 139s to find the best ballot-polling and comparison audit, respectively.

In 22/26 of the elections in Table \ref{tab:AUDITIRV-NOERRORS}, the ballot-polling audit generated by RAIRE required a similar number of ballot samples to that of the best alternate method (EO, SE, or WO). In the remaining 4 instances, the RAIRE audit was significantly more efficient. Consider instances 12 and 13 -- the Oakland 2012 City Council election for District 3, and the Pierce 2008 County Assessor election. For these instances, neither the EO, SE, or WO methods were able to avoid a full recount. The RAIRE audits were able to confirm the reported outcomes in these elections by sampling no more than 17\% of the cast ballots, on average. The comparison audits generated by RAIRE are significantly more efficient than their ballot-polling counterparts, across our suite of election instances. 

Consider instance 17 in Table \ref{tab:AUDITIRV-NOERRORS} -- the Aspen 2009 City Council election. We can, with a RAIRE comparison audit, confirm the reported outcome (with risk limit $\alpha = 5$\%) by sampling just under 10\% of the cast ballots (254 ballots), on average. If we were to isnstead use one of the EO, SE, or WO approaches of conducting a comparison audit, we would need to sample just under 83\% of the cast ballots (2112 ballots), on average. Table \ref{tab:AUDITIRV-NOERRORS} also shows that as the $\gamma$ parameter increases, the number of ballots checked in a comparison audit may increase slightly, but not significantly. This parameter has more influence on selecting an initial sample size when sampling ballots in batches \cite{stark2010super}.

We have shown that RAIRE is able to find efficient ballot-polling and ballot-level comparison audit configurations across a range of example elections, in the context where electronic ballot records exactly match their corresponding paper ballot. In Appendix \ref{app:WithErrors}, we consider the effectiveness of our audits in the setting where varying numbers of errors (or discrepancies) are introduced into the reported (electronic or digitised) ballot records.  We show that even when there are discrepancies between actual and reported ballots: comparison audits are still more efficient, in general, than their ballot-polling counterparts; and RAIRE is able to generate efficient audits that sample only a small fraction of cast ballots.

\section{Conclusion}

We have presented and evaluated several methods for conducting ballot-polling and ballot-level comparison RLAs for IRV elections. These approaches represent the first practical techniques for conducting RLAs for IRV. As in FPTP, we find that comparison-based IRV audits are, in general, more efficient than their ballot-polling counterparts. These audits typically require only a small fraction of cast ballots to be sampled, though very close elections (with a MOV that is less than 1\% of cast ballots, for example) generally require a full manual recount. We have presented an algorithm, RAIRE, for designing efficient ballot-polling and ballot-level comparison RLAs for a given IRV election. This algorithm finds a collection of assertions to audit that require the least number of expected ballot checks to confirm (assuming the announced outcome is correct), while still guaranteeing that a wrong result with be detected with a probability of at least $1-\alpha$. The audit configurations generated by RAIRE are competitive with alternate methods considered in this paper, and in some cases are substantially more efficient.

\afterpage{
\begin{landscape}
\begin{table}[!t]
\footnotesize
\centering
\begin{tabular}{|l|l||cc|cc||cc|cc|cc|cc|}
\multicolumn{14}{r}{\textbf{Auditing using RAIRE via ballot-polling (BP) and comparison (CP) audits, $\alpha = 5$\%}} \\
\hline
	&    & \multicolumn{4}{c||}{BP}       & \multicolumn{8}{c|}{CP} \\
\cline{3-14}
	&    &  \multicolumn{2}{c|}{Best Alt.} & \multicolumn{2}{c||}{ RAIRE } & \multicolumn{2}{c|}{Best Alt. $\gamma = 1.1$} & \multicolumn{2}{c|}{$\gamma = 1.1$ } & \multicolumn{2}{c|}{$\gamma = 1.2$} & \multicolumn{2}{c|}{$\gamma = 1.3$ } \\
	\textbf{\#} & $|\mathcal{B}|$ & \multicolumn{2}{c|}{{\scriptsize EO/SE/WO}} &            &         &  \multicolumn{2}{c|}{{\scriptsize EO/SE/WO}}   & \multicolumn{2}{c|}{RAIRE} &   \multicolumn{2}{c|}{RAIRE}   &  \multicolumn{2}{c|}{RAIRE} \\
	&     & Method   &  Polls \%             &  Polls \%  & ASN \%  & Method & Polls \% &  Polls \%  & ASN \% & Polls \% & ASN \%  & Polls \%  & ASN \%     \\        
\hline 
	1  &4,682 &EO & 3.9 &  5.4 & 4.7    & \bf SE & \bf 0.9  & \bf 0.9 & \bf 0.9 & 1& 1     & 1.1& 1.1\\
2  & 5,333&WO & 0.8 & 0.9 & 0.9     & \bf WO & \bf 0.4  & \bf 0.4 & \bf 0.4 & \bf 0.4& \bf 0.4 & 0.4&0.5 \\
	3  &14,040 &WO & 0.3& 0.3 & 0.3     & \bf  WO & \bf 0.1 &\bf 0.1 &\bf 0.1  & \bf 0.1& \bf 0.1 & \bf 0.1& \bf 0.2 \\
	4  &43,661 &EO,SE&  1.8& 1.5 & 1.4     & \bf SE,EO & \bf 0.2 & \bf 0.2 & \bf 0.2  &\bf 0.2& \bf 0.2 & \bf 0.2& \bf 0.2\\
	5  &159,987 &EO,SE&  0.2& 0.3 & 0.3     & \bf SE & \bf 0.04  & \bf 0.04 &\bf 0.04  &\bf  0.04&\bf 0.04 & 0.1& 0.1\\
	6  &2,544 &EO&  52.7& 28.1 & 46.9   & \bf SE & \bf 3.7 &\bf 3.7 & \bf 3.8 & 4& 4.1 & 4.4& 4.4\\ 
	7  &6,426 &WO&  0.8& 0.6 & 0.6     & \bf WO  &\bf  0.3 &\bf 0.3 & \bf 0.3 & \bf 0.3& \bf 0.3 &\bf 0.3 & \bf 0.3\\
	8  &5,708 &WO&  3.8& 1.6 & 2.7     & WO &  0.9 &\bf 0.6 & \bf 0.7 & 0.7& 0.7 & 0.7& 0.8\\
	9  &13,482 &EO&  7.3& 5.2 & 6.7     & \bf SE  & \bf 0.7 & \bf 0.7 & \bf 0.7 & \bf 0.7& \bf 0.7 & 0.8& 0.8 \\ 
	10 &312,771 &EO,SE&  7.6& 13.9 & 9.8    & \bf SE,EO & \bf  0.2 & \bf 0.2 &\bf  0.2 &\bf  0.2& \bf 0.2 & \bf 0.2& \bf 0.2  \\
	11 &28,703 &WO&  0.8& 0.8 & 0.6      & WO &  0.3 & \bf 0.1 &\bf  0.1 & 0.2& 0.2 & 0.2& 0.2\\
	12 &26,761 &--&  $\infty$ & 14.2 & 13.1   & SE,EO  &  14.6 &\bf 0.9 & \bf 0.9 & \bf 0.9& \bf 0.9 & 1& 1   \\
	13 &312,771 &--&  $\infty$ & 17 & 22.7     & SE,EO & 2.4  & \bf 0.3 & \bf 0.3 & \bf 0.3&  \bf 0.3& 0.4& 0.4 \\
	14 &23,494 &EO,SE & 92.9 & 87.6 & $\infty$& \bf SE & \bf 2.8 & \bf 2.8 & \bf 2.8 & 3.1& 3.1 & 3.4& 3.4 \\
	15 &57,492 &WO&  0.1& 0.1 & 0.1      & WO & 0.1 &\bf 0.04 &\bf 0.04  &\bf  0.04& \bf 0.04 & \bf 0.04& \bf 0.1\\
	16 &23,884 &WO&  0.6& 0.6 & 0.5      & WO & 0.3 & \bf 0.1 & \bf 0.1 & 0.2&  0.2& 0.2& 0.2\\ 
	17 &2,544 &--&  $\infty$ & $\infty$&$\infty$& SE& 82.4& \bf 9.4 & \bf 9.5 & 10.2& 10.3 & 11.1& 11.1\\
	18 &122,268 &SE&  15  & 15.3 & 15.5   & SE &  0.4 & \bf 0.3 & \bf 0.3 & 0.4&  0.4  & 0.4& 0.4\\
	19 &101,431 &WO&  0.5 & 5.4 & 0.1     & WO &  0.4 & \bf 0.04 & \bf 0.04 & \bf 0.04 &  \bf 0.04& \bf 0.04& \bf 0.04\\
	20 &149,465 &\bf WO&  \bf 0.01& \bf 0.01 &\bf  0.01   & \bf SE,WO &\bf  0.01  &\bf 0.01 &\bf  0.01 & \bf 0.01& \bf 0.01 & \bf 0.01& \bf 0.01\\
	21 &79,415 &WO&  0.3& 0.2 & 0.2     & WO  & 0.2 & \bf 0.1 & \bf 0.1 & \bf 0.1& \bf 0.1 & \bf 0.1& \bf 0.1 \\
\hline
	22 &46,952 &WO&  3.7 & 3.2 & 1.9     & WO & 0.7  & \bf 0.2& \bf 0.2 &\bf 0.2 &\bf 0.2 &\bf 0.2&\bf 0.2 \\
	23 &45,124 &WO&  0.9& 0.8 & 0.7     & \bf WO & \bf 0.1 &\bf 0.1&\bf 0.1 &\bf 0.1 &\bf 0.1 &\bf 0.1 &\bf  0.1\\
	24 &48,259 &--&  $\infty$ & $\infty$ & $\infty$& \bf SE& \bf 3.2&  \bf 3.2& \bf 3.2 &3.5 &3.5 &3.8 & 3.8 \\
	25 &47,698 &WO&  0.5& 0.5 &0.3      & \bf WO & \bf 0.1  &\bf 0.1&\bf 0.1 &\bf 0.1 &\bf 0.1& \bf 0.1  & \bf 0.1\\
	26 &42,747 &SE&  1& 1.3 & 0.7     &  SE & 0.3 & \bf 0.1& \bf 0.1 & \bf 0.1& \bf 0.1& \bf 0.1& \bf 0.1 \\
\hline
\end{tabular}
\caption{Comparison of the best ballot-polling and ballot-level comparison audit methods across 26 IRV elections. We compare the average number of ballot samples required (expressed as a percentage of total ballots cast) by the best alternate ballot-polling or comparison audit methods (EO, SE, and WO) and those generated by RAIRE. The notation $\infty$ indicates a percentage of ballots (or ASN) greater than 100\%. }
\label{tab:AUDITIRV-NOERRORS}
\end{table}
\end{landscape}}

\bibliographystyle{spbasic}
\bibliography{BIB}

\appendix

\section{RAIRE in elections with discrepancies}\label{app:WithErrors}
We introduce discrepancies between reported and actual ballots according to a defined error rate, which we vary between 1\% and 5\%. This means that for any given ballot, there is a 1\% to 5\% probability that its electronic version differs, in some way, from the paper version. The electronic record of a paper ballot is a partial or complete sequence of candidates, ordered according to voter preference.
 We introduce an error in a reported ballot record with one of the following operations: replacing a randomly selected candidate in this preference ordering with a randomly selected candidate that does not appear in the ordering; inserting a randomly selected candidate that does not appear in the ordering into a randomly selected position; flipping the positions of two randomly selected candidates in the ordering; and removing a randomly selected candidate in the ordering. For each reported ballot, we introduce an error with a probability equal to the error rate. When introducing an error, we uniformly randomly choose one of the above manipulations to perform.

\afterpage{
\begin{landscape}
\begin{table}[!t]
\footnotesize
\centering
\begin{tabular}{|l|l|cc|cc||cc|cc|cc|cc|}
\multicolumn{14}{r}{\textbf{Auditing using RAIRE via ballot polling (BP) and comparison (CP) audits, $\alpha = 5$\%, 1\% errors }} \\
\hline
	&    & \multicolumn{4}{c||}{BP}       & \multicolumn{8}{c|}{CP} \\
\cline{3-14}
	&     &  \multicolumn{2}{c|}{Best Alt.} & \multicolumn{2}{c||}{ RAIRE } & \multicolumn{2}{c|}{Best Alt. $\gamma = 1.1$} & \multicolumn{2}{c|}{$\gamma = 1.1$ } & \multicolumn{2}{c|}{$\gamma = 1.2$} & \multicolumn{2}{c|}{$\gamma = 1.3$ } \\
	\textbf{\#} & $|\mathcal{B}|$ & \multicolumn{2}{c|}{{\scriptsize EO/SE/WO}} &            &         &  \multicolumn{2}{c|}{{\scriptsize EO/SE/WO}}   & \multicolumn{2}{c|}{RAIRE} &   \multicolumn{2}{c|}{RAIRE}   &  \multicolumn{2}{c|}{RAIRE} \\
	&      & Method   &  Polls \%             &  Polls \%  & ASN \%  & Method & Polls \% &  Polls \%  & ASN \% & Polls \% & ASN \%  & Polls \%  & ASN \%     \\        
\hline 
	1  &4,682 &EO & 3.7 &  3.8 &  4.8                 &\bf SE    & \bf 0.9     &\bf 0.9 &\bf 0.9  & 1& 1     &1.1 & 1.1\\
	2  &5,333 &WO & 0.9 &  0.9 &  0.9                 &\bf  WO    & \bf 0.4     & \bf 0.4& \bf 0.4 & \bf 0.4& \bf 0.4    & \bf 0.4& \bf 0.5\\
	3  &14,040 &WO & 0.2 &  0.2 &  0.3                 & \bf WO    & \bf 0.1     &\bf  0.1&\bf 0.1 &\bf 0.1&\bf 0.1     &0.2 & 0.2 \\
	4  &43,661 &EO & 1.2 &  1.3 &  1.5                 & \bf SE,EO &\bf 0.2     &\bf 0.2&\bf 0.2 &\bf 0.2&\bf 0.2     &\bf 0.2&\bf 0.2\\
	5  &159,987 &EO & 0.4 &  0.4 &  0.3                 & \bf SE    &\bf 0.04    &\bf 0.04&\bf 0.04 &0.04 &  0.04    & 0.1& 0.1\\
	6  &2,544 &EO & 36.2 & 36.2  &  48                & \bf SE    & \bf 3.8     & \bf 3.8& \bf 3.8 & 4.3&  4.1    & 4.6& 4.5\\ 
	7  &6,426 &WO & 0.6 &  0.6 &  0.6                 & \bf WO    & \bf 0.3     &\bf 0.3&\bf 0.3 &\bf 0.3&\bf  0.3    & \bf 0.3& \bf 0.4\\
	8  &5,708 &WO & 1.3 &  1.5 & 2.7                  & WO    & 0.9     & \bf 0.6& \bf 0.7 & 0.7&  0.7    &0.7 & 0.8\\
	9  &13,482 &EO & 4.5 & 6.2  & 6.7                  & SE    & 0.9     & \bf 0.7& \bf 0.7 & 0.8&  0.7    & 0.8& 0.8\\ 
	10 &312,771 &EO & 16.8 & 17.8  & 9.1                & \bf SE,EO & \bf 0.2     & \bf 0.2& \bf 0.2 & \bf 0.2&  \bf 0.2    &\bf 0.2& \bf 0.2  \\
	11 &28,703 &WO &0.9  &  0.7 &   0.6                & WO    & 0.3     & \bf 0.1& \bf 0.2 & 0.2& 0.2    & 0.2&0.2 \\
	12 &26,761& -- & $\infty$ & 12.3  &  13.2          & SE,EO & 62.4    & \bf 0.9& \bf  0.9& \bf 0.9& \bf 0.9    & 1& 1  \\
	13 &312,771& -- & $\infty$ & 12.9 &  24.1           & --    & $\infty$& \bf 0.4& \bf 0.3 &\bf 0.4&\bf  0.3    &\bf 0.4&\bf 0.4 \\
	14 &23,494& EO & 87 &  90 &  $\infty$              & \bf SE    &\bf 4.3     &\bf 4.3&\bf 2.8 &\bf 4.3&\bf  3    &\bf 4.3&\bf 3.3 \\
	15 &57,492& WO & 0.2 & 0.2  &   0.1                & WO    & 0.1     & \bf 0.04&\bf 0.04 &\bf 0.04 &\bf 0.04 &\bf 0.04&\bf 0.1\\
	16 &23,884& WO & 0.5 &  0.6 &  0.5                 & WO    & 0.3     &\bf 0.1 & \bf 0.1 & 0.2&  0.2    & 0.2& 0.2\\ 
	17 &2,544& -- & $\infty$ & $\infty$  &  $\infty$  & SE    & 98.6    &\bf 12.6 & \bf 9.6 & 13.1& 10.5     & 13.7& 11.3\\
	18 &122,268& SE & 26.3 & 16.3  &  14.9              & SE    & 0.6     &\bf 0.4 &\bf 0.3 &\bf 0.4&\bf  0.4    &\bf 0.4 &\bf 0.4\\
	19 &101,431& WO & 0.6 & 0.3  &  0.1                 & WO    & 0.5     &\bf 0.04 & \bf 0.04 &\bf 0.04&\bf  0.04    &\bf 0.04 &\bf  0.04\\
	20 &149,465& \bf WO & \bf 0.01 & \bf 0.01 & \bf 0.01& \bf SE,WO &\bf 0.01    &\bf 0.01 &\bf 0.01 &\bf 0.01&\bf  0.01    &\bf 0.01&\bf 0.01 \\
	21 &79,415& WO& 0.3 &  0.2 &  0.2                  & WO    & 0.2     &\bf 0.1 &\bf 0.1 &\bf 0.1&\bf  0.1    &\bf 0.1 &\bf 0.1 \\
\hline
	22 &46,952& WO & 3.9     & 2.1  &  1.9           & WO & 0.9  & \bf 0.2& \bf 0.2  &\bf 0.2 & \bf 0.2 &\bf 0.2&\bf 0.2 \\
	23 &45,124& WO & 0.7     & 0.7  &  0.7           & WO & 0.2  &\bf 0.1&\bf 0.1   &\bf 0.1 & \bf 0.1 &\bf 0.1 &\bf 0.1 \\
	24 &48,259& -- & $\infty$& $\infty$  &  $\infty$ & SE & 16.3 & 16.3& 2.9  &11.4 &  3.1 & \bf 10& \bf 3.4  \\
	25 &47,698& WO & 0.3     & 0.4  &  0.3           & \bf WO & \bf 0.1  &\bf 0.1&\bf 0.1   &\bf 0.1 &\bf  0.1 &\bf 0.1 &\bf 0.1  \\
	26 &42,747& WO& 1.2      & 1.2  &  0.7           & SE & 0.3  & \bf 0.1&\bf 0.1   &\bf 0.1 &\bf  0.1 &\bf 0.1 &\bf 0.1 \\
\hline
\end{tabular}
\caption{Comparison of the best ballot-polling and ballot-level comparison audit methods across 26 IRV elections, with an error rate of 1\% used to manipulate reported ballots. The average \# of ballot samples required (expressed as a percentage of ballots cast) by the best alternate method (EO, SE, and WO) and those generated by RAIRE are compared. The notation $\infty$ indicates a percentage of ballots (or ASN) greater than 100\%.  }
\label{tab:AUDITIRV-1PCERRORS}
\end{table}
\end{landscape}}

In this setting, we simulate each auditing approach 50 times -- with 10 different seeds used to inject errors into electronic (reported) ballot records, and 5 seeds used to randomly draw (sample) ballots during the audit. When reporting the ASNs and actual number of ballots sampled by each auditing method, we average these values over the 50 simulated audits. Tables~\ref{tab:AUDITIRV-1PCERRORS} to~\ref{tab:AUDITIRV-5PCERRORS} report the ASN and actual number of ballot samples required, on average, across the simulation of varying types of audit in each of our 26 elections, with a 1\% to 5\% error rate, $\alpha = 5$\%, and $\gamma = \{1.1 \ldots 1.3\}$. We compare the EO, SE, and WO auditing methods, in both a ballot-polling and comparison audit context, against the audits generated by RAIRE. 

Tables~\ref{tab:AUDITIRV-1PCERRORS} to~\ref{tab:AUDITIRV-5PCERRORS} show that even when there are discrepancies between actual and reported ballots: comparison audits are still more efficient, in general; and RAIRE is able to generate efficient audits that sample only a small fraction of cast ballots. Across Tables~\ref{tab:AUDITIRV-1PCERRORS} to~\ref{tab:AUDITIRV-5PCERRORS}, we bold the audit with the lowest average number of ballot polls required when simulated (expressed as a percentage of total ballots cast). 

As the rate of introduced errors increases toward 5\%, the ASNs associated with the comparison audits generated by RAIRE significantly underestimate the actual auditing effort required in a small number of instances. This is the case in instances 10 (Pierce 2008 County Executive), 13 (Pierce 2008 County Assessor), 14 (San Leandro 2010 Mayor), 17 (Aspen 2009 City Council), 18 (Oakland 2010 Mayor) and 24 (Gosford NSW 2015). The MOV in each of these elections is less than 1\% of the total ballots cast. 

Our results indicate that for very close elections, with a very small margin of victory, the impact of each discrepancy encountered in the sampling of ballots has a significant influence on the statistics being maintained throughout the comparison audit. Recall that the MACRO algorithm of Figure~\ref{fig:macro} repeatedly samples ballots until a running Kaplan-Markov MACRO
P-value ($P_{KM}$) falls below the given risk limit $\alpha$. When we discover a discrepancy that has resulted in the margin between a winning and losing candidate being overstated (i.e., thought to be larger than it actually is), this $P_{KM}$ statistic increases at a rate that is proportional to the inverse of the election MOV. For elections with a very small MOV, each discovered error may significantly increase the ASN of the audit. In these instances, a full manual recount is likely to be required (and indeed, the announced outcome may be wrong).    

\afterpage{
\begin{landscape}
\begin{table}[!t]
\footnotesize
\centering
\begin{tabular}{|l|l|cc|cc||cc|cc|cc|cc|}
\multicolumn{14}{r}{\textbf{Auditing using RAIRE via ballot polling (BP) and comparison (CP) audits, $\alpha = 5$\%, 3\% errors }} \\
\hline
	&   & \multicolumn{4}{c||}{BP}       & \multicolumn{8}{c|}{CP} \\
\cline{3-14}
	&    &  \multicolumn{2}{c|}{Best Alt.} & \multicolumn{2}{c||}{ RAIRE } & \multicolumn{2}{c|}{Best Alt. $\gamma = 1.1$} & \multicolumn{2}{c|}{$\gamma = 1.1$ } & \multicolumn{2}{c|}{$\gamma = 1.2$} & \multicolumn{2}{c|}{$\gamma = 1.3$ } \\
	\textbf{\#} & $|\mathcal{B}|$ & \multicolumn{2}{c|}{{\scriptsize EO/SE/WO}} &            &         &  \multicolumn{2}{c|}{{\scriptsize EO/SE/WO}}   & \multicolumn{2}{c|}{RAIRE} &   \multicolumn{2}{c|}{RAIRE}   &  \multicolumn{2}{c|}{RAIRE} \\
	&      & Method   &  Polls \%             &  Polls \%  & ASN \%  & Method & Polls \% &  Polls \%  & ASN \% & Polls \% & ASN \%  & Polls \%  & ASN \%     \\        
\hline 
	1  & 4,682&SE & 3.7     & 3.8  &  5            &\bf  SE & \bf 1        & \bf 1 & \bf 0.9    & 1.1& 1&  1.1   & 1\\
	2  & 5,333&WO & 0.9     & 0.9  &  0.9          & WO & 0.5      &\bf 0.4 &\bf 0.4   &\bf 0.4&\bf 0.4&\bf  0.4   &\bf 0.4\\
	3  & 14,040&WO & 0.2     & 0.2  &  0.3          & \bf WO & \bf 0.1      & \bf 0.1 & \bf 0.1  &\bf 0.1 & \bf 0.2& \bf 0.1   &\bf 0.2\\
	4  & 43,661&SE & 1.2     & 1.3  &  1.6          & \bf SE,EO &\bf  0.2  &\bf 0.2 &\bf 0.2  &\bf 0.2&\bf 0.2&\bf  0.2   &\bf 0.2\\
	5  & 159,987&SE,EO & 0.4  & 0.4  &  0.3          & \bf SE & 0.04     &\bf 0.04 &\bf 0.04& 0.1& 0.04&  0.1   & 0.1\\
	6  & 2,544&EO & 36.2    & 36.2 & 49.7          & \bf SE & \bf 3.9      & \bf 3.9 & \bf 3.8  & 4.4& 4.2&  4.7   & 4.6\\ 
	7  & 6,426&WO & 0.6     & 0.6  &  0.7          & \bf WO &\bf 0.3      &\bf 0.3 &\bf 0.3  &\bf 0.3 &\bf 0.3&\bf  0.3   &\bf 0.4\\
	8  & 5,708&WO & 1.3     & 1.5  &  2.8          & WO & 0.9      & \bf 0.7 & \bf 0.7  &\bf 0.7 &\bf 0.7&  0.8   & 0.8\\
	9  & 13,482&EO & 4.5     & 6.2  &  6.9          & SE &  0.9     & \bf 0.8 & \bf 0.7  &\bf 0.8 &\bf 0.8&  0.9   & 0.8\\ 
	10 & 312,771&SE,EO & 18   & 18.2 &  8.1          & SE,EO & 0.7   & 0.7 & 0.2  &\bf 0.5 &\bf 0.2& \bf 0.5   &\bf 0.2  \\
	11 & 28,703&WO & 0.9     & 0.7  &   0.6         & WO & 0.4      & \bf 0.2 & \bf 0.2  &\bf 0.2 &\bf 0.2& \bf 0.2   &\bf 0.2\\
	12 & 26,761&-- & $\infty$& 12.3 &  12.6         & EO,SE & $\infty$& \bf 1.2 &\bf 0.8& \bf 1.2&\bf 0.9&\bf  1.2     & \bf 1  \\
	13 & 312,771&-- & $\infty$& 13.2 &  27.2         & EO,SE & $\infty$& 1 & 0.3  &0.8&0.4&\bf 0.7     &\bf 0.4 \\
	14 & 23,494&SE,EO & 87.1 & 90   &  $\infty$     & SE & 19.1     & 19.1 & 2.8 & 11.8 &3.1&\bf 9.2   &\bf 3.3 \\
	15 & 57,492&WO & 0.2     & 0.2  &  0.1          & WO & 0.1      &\bf 0.04 &\bf 0.04&\bf 0.04 &\bf 0.04&  0.1   & 0.1\\
	16 & 23,884&WO & 0.5     & 0.6  &  0.5          & WO & 0.4      & \bf 0.2 &\bf 0.2  &\bf 0.2 &\bf 0.2&\bf  0.2   &\bf 0.2\\ 	
	17 & 2,544&--&$\infty$   &$\infty$& $\infty$    & -- & $\infty$& 17.6 &\bf 9.7   & \bf 16.6 & 10.5& 16.8   & 11.4\\
	18 & 122,268& SE & 16.4    & 16.4 &  13.2         & SE & 19.5    & 0.8 & 0.3   & \bf 0.6 &\bf 0.3&\bf 0.6   &\bf 0.4\\
	19 & 101,431&-- & $\infty$& 0.3  &   0.1         & WO &  0.6     & \bf 0.04 &\bf 0.04&\bf 0.04 &\bf 0.04& 0.1    &0.04 \\
	20 & 149,465&\bf WO &\bf 0.01    &\bf 0.01 &\bf  0.01 & \bf SE,WO &\bf 0.01  &\bf 0.01 &\bf 0.01&\bf 0.01 &\bf 0.01&\bf 0.01 &\bf 0.01\\
	21 &79,415 & WO & 0.3     & 0.2  &  0.2          & WO & 0.2     &\bf 0.1 &\bf 0.1   & \bf 0.1  &\bf 0.1& \bf 0.1   &\bf 0.1\\
\hline
	22 &46,952& WO & 4.3     & 2.1  &  1.9          & WO & 1.2  &\bf 0.2 &\bf 0.2 &\bf 0.2 &\bf  0.2    &\bf 0.2&\bf 0.2 \\
	23 &45,124& SE,WO & 0.7  & 0.7  &  0.7          & \bf SE & \bf 0.1 & 0.2 & 0.2 & \bf 0.1 & \bf 0.1    &\bf 0.1&\bf 0.1 \\
	24 &48,259&--&$\infty$&$\infty$ & $\infty$      & SE & 76.8 &76.8 & 2.3 &69 &  2.5    &\bf 62.3 &\bf 2.7  \\
	25 &47,698& WO & 0.3     & 0.4  &  0.4          & WO & 0.2  & \bf 0.1 &\bf 0.1 &\bf 0.1 &\bf 0.1    &\bf 0.1 &\bf 0.1  \\
	26 &42,747& WO& 1.2      & 1.2  &  0.7          & WO & 0.4  & \bf 0.1& \bf 0.1 &\bf 0.1 &\bf 0.1    &\bf 0.1 &\bf 0.1 \\
\hline
\end{tabular}
\caption{Comparison of the best ballot-polling and ballot-level comparison audit methods across 26 IRV elections, with an error rate of 3\% used to manipulate reported ballots. The average \# of ballot samples required (expressed as a percentage of ballots cast) by the best alternate method (EO, SE, and WO) and those generated by RAIRE are compared. The notation $\infty$ indicates a percentage of ballots (or ASN) greater than 100\%.  }
\label{tab:AUDITIRV-3PCERRORS}
\end{table}
\end{landscape}}

\afterpage{
\begin{landscape}
\begin{table}[!t]
\footnotesize
\centering
\begin{tabular}{|l|l|cc|cc||cc|cc|cc|cc|}
\multicolumn{14}{r}{\textbf{Auditing using RAIRE via ballot polling (BP) and comparison (CP) audits, $\alpha = 5$\%, 5\% errors }} \\
\hline
	&    & \multicolumn{4}{c||}{BP}       & \multicolumn{8}{c|}{CP} \\
\cline{3-14}
	&  &  \multicolumn{2}{c|}{Best Alt.} & \multicolumn{2}{c||}{ RAIRE } & \multicolumn{2}{c|}{Best Alt. $\gamma = 1.1$} & \multicolumn{2}{c|}{$\gamma = 1.1$ } & \multicolumn{2}{c|}{$\gamma = 1.2$} & \multicolumn{2}{c|}{$\gamma = 1.3$ } \\
	\textbf{\#} & $|\mathcal{B}|$ & \multicolumn{2}{c|}{{\scriptsize EO/SE/WO}} &            &         &  \multicolumn{2}{c|}{{\scriptsize EO/SE/WO}}   & \multicolumn{2}{c|}{RAIRE} &   \multicolumn{2}{c|}{RAIRE}   &  \multicolumn{2}{c|}{RAIRE} \\
	&    & Method   &  Polls \%             &  Polls \%  & ASN \%  & Method & Polls \% &  Polls \%  & ASN \% & Polls \% & ASN \%  & Polls \%  & ASN \%     \\        
\hline 
	1  & 4,682&SE,EO & 3.7 & 3.7  &  5          & \bf SE &\bf  1.1     &\bf 1.1 &\bf 1    &\bf 1.1& \bf 1    & 1.2& 1.1\\
	2  & 5,333&WO & 0.9    & 0.9  &  1          & WO &  0.5     & \bf 0.4 & \bf 0.4  & \bf 0.4& \bf 0.4    & \bf 0.4& \bf 0.5 \\
	3  & 14,040&WO & 0.2    & 0.2  &  0.3        & \bf WO &\bf  0.1  &\bf 0.1 &\bf 0.1  &\bf 0.1&\bf  0.2    & 0.2& 0.2 \\
	4  & 43,661&SE,EO & 1.3 & 1.3  &  1.7        & EO &  0.3     &\bf 0.2 &\bf 0.2  &\bf 0.2&\bf  0.2    & 0.3& 0.2\\
	5  & 159,987&SE,EO &  0.4& 0.4  &  0.3        & \bf SE & \bf 0.1  &\bf 0.1 &\bf 0.04 &\bf 0.1&  \bf 0.04    &\bf  0.1&\bf 0.1\\
	6  & 2,544&EO & 36.3   & 36.3  &  50.7      & SE &  5.5     &5.5 & 3.9  &\bf 5.4&\bf 4.3     & 5.6& 4.6\\ 
	7  & 6,426&WO &  0.6   & 0.6  &  0.7        & \bf WO & \bf 0.3     &\bf 0.3 &\bf 0.3  &\bf 0.3&\bf 0.3     &\bf 0.3&\bf 0.4\\
	8  & 5,708&WO & 1.4    & 1.5  &  2.9        & WO &  1       &\bf 0.7 &\bf 0.7  & 0.8& 0.7     & 0.8& 0.8\\
	9  & 13,482&EO & 4.5    & 6.2  &  7          & SE &  1.1     &\bf 0.9 &\bf 0.7  &\bf 0.9&\bf 0.8     &\bf 0.9&\bf 0.8\\ 
	10 & 312,771&SE,EO & 18.5 & 18.5  &  7.1      & SE,EO & 18.8  &19.7& 0.1  & \bf 1.4&\bf 0.2     &\bf 0.7&\bf  0.2 \\
	11 & 28,703&WO & 0.9     &  0.7 & 0.6        & WO &  0.4     & \bf 0.2 &\bf 0.2  &\bf 0.2&\bf 0.2     &\bf 0.2&\bf 0.2\\
	12 & 26,761&-- & $\infty$ & 12.3  &  13.2    & -- & $\infty$ & \bf 1.3 &\bf 0.9  &\bf 1.3&\bf 0.9     &\bf 1.3&\bf 1  \\
	13 & 312,771&-- & $\infty$ &  13.3 &  30      & -- & $\infty$ &27.1 & 0.3 &\bf 2.7&\bf 0.4     &\bf 1.4&\bf 0.4 \\
	14 & 23,494&SE & 86.3     &  89.5 &  $\infty$& SE & 58.8     &58.8 & 2.6 & 40.5& 2.8     &\bf 26.4&\bf 3 \\
	15 & 57,492&WO &  0.2     & 0.2  &  0.1      & WO & 0.1      &\bf 0.04 &\bf 0.04 & 0.1&  0.1    & 0.1& 0.1\\
	16 & 23,884&WO &  0.5   &  0.7   & 0.5     & WO  & 0.4     &\bf 0.2 &\bf  0.2  &\bf  0.2  &\bf 0.2  &\bf  0.2& \bf 0.2\\ 
	17 & 2,544&-- &  $\infty$&  $\infty$ & $\infty$&--&$\infty$ &22.1 &10.2 & 19.6 & 11.1& \bf 19.3&\bf 12.1\\
	18 & 122,268&SE & 25.7     & 16.5  &  12.1    & -- & $\infty$ &2.6 & 0.3  & 1.1&  0.3    &\bf 0.9&\bf 0.4\\
	19 & 101,431&WO & 0.7      & 0.3  &  0.1      & WO &  0.7     &\bf 0.04 &\bf 0.04&\bf 0.04&\bf  0.04    & 0.1& 0.04\\
	20 & 149,465&\bf WO&\bf 0.01     &\bf 0.01  & \bf 0.01    &\bf SE,WO &\bf 0.01  &\bf 0.01 &\bf 0.01&\bf 0.01&\bf  0.01 &\bf 0.01&\bf 0.01 \\
	21 & 79,415 &WO& 0.3       & 0.2  &  0.2      & WO &  0.2     &\bf 0.1 &\bf 0.1  &\bf 0.1& \bf  0.1   &\bf 0.1 &\bf 0.1 \\
\hline
	22 & 46,952 &SE & 3.2 & 2.1  &  1.9      & \bf SE & \bf 0.3  &\bf 0.3 &\bf 0.2   &\bf 0.3&\bf  0.2   &\bf 0.3 &\bf 0.2 \\
	23 & 45,124&SE & 0.7 &  0.7 &  0.7      & \bf SE & \bf 0.1  &\bf 0.1 &\bf  0.1   &\bf 0.1&\bf  0.1   &\bf 0.1 &\bf 0.1 \\
	24 & 48,259&SE & 97.9 & 97.9  & $\infty$& SE & 90.1  &90.1 & 1.9 & 92&   2    & \bf 87 &\bf 2.2  \\
	25 & 47,698&WO & 0.3 &  0.4 &  0.4      & WO &  0.2 & \bf 0.08&  \bf 0.08&0.1 &  0.1   &0.1 & 0.1  \\
	26 & 42,747&WO& 1.2 & 1.2  & 0.7        & WO &  0.5 & \bf 0.1& \bf 0.1   &\bf 0.1 & \bf 0.1   &0.2 & 0.1 \\
\hline
\end{tabular}
\caption{Comparison of the best ballot-polling and ballot-level comparison audit methods across 26 IRV elections, with an error rate of 5\% used to manipulate reported ballots. The average \# of ballot samples required (expressed as a percentage of ballots cast) by the best alternate method (EO, SE, and WO) and those generated by RAIRE are compared. The notation $\infty$ indicates a percentage of ballots (or ASN) greater than 100\%. }
\label{tab:AUDITIRV-5PCERRORS}
\end{table}
\end{landscape}}

\end{document}